\documentstyle[aps,eqsecnum,epsf,prd]{revtex}
\begin{document}
\preprint{\vbox{
\hbox{UCSD/PTH 95--19}
\hbox{hep-ph/9511233}
}}
\title{Semileptonic $B$ and $\Lambda_b$ Decays  and
Local Duality in QCD}
\author{C. Glenn Boyd, Benjamin Grinstein and Aneesh V.~Manohar}
\address{Department of Physics 0319, University of California at San Diego,\\
9500 Gilman Drive, La Jolla, CA 92093-0319}
\date{November 1995}
\maketitle
\widetext
\vskip-1.50in
\rightline{\vbox{
\hbox{UCSD/PTH 95--19}
\hbox{hep-ph/9511233}
}}
\vskip1.5in

\begin{abstract}
The inclusive and exclusive semileptonic decay distributions for
$b\rightarrow c$ decay are computed in the Shifman-Voloshin limit. The
inclusive decay distributions (computed using an operator product
expansion) depend on quark masses, and the exclusive decay
distributions depend on hadron masses. Nevertheless, we show
explicitly how the first two terms in the $1/m$ expansion match
between the inclusive and exclusive decays. Agreement between the
inclusive and exclusive decay rates requires a minimum smearing region
of size $\Lambda_{\rm QCD}$ before local duality holds in QCD.  The
$\alpha_s$ corrections to the inclusive and exclusive decay rates are
also shown to agree to order $(\log m)/m^2$.  The $\alpha_s/m^2$
corrections are used to obtain the $\alpha_s$ correction to Bjorken's
inequality on the slope of the Isgur-Wise function.
\end{abstract}



\section{Introduction}\label{sec:intro}

The semileptonic decay of hadrons containing $b$ quarks allows one to
measure the $V_{ub}$ and $V_{cb}$ elements of the quark mixing
matrix. Reliable model independent values for these matrix elements
can only be obtained if one can accurately calculate the hadronic
decay distributions (or decay rates) from QCD. Two approaches to
semileptonic $b\to c$ decay are to study exclusive decay modes such as
$B\rightarrow De^-\nu_e$, $B\rightarrow D^*e^-\nu_e$, or
$\Lambda_b\rightarrow \Lambda_ce^-\nu_e$, or inclusive decay modes
such as $B\rightarrow X_{u,c} e^-\nu_e$. The inclusive and exclusive
modes have been analyzed using heavy quark effective theory
(HQET)~\cite{hqet,SV}.

The exclusive decay rates can be obtained by computing decay form
factors using HQET, and then integrating over the allowed phase
space. The final answer typically depends on the hadron masses, and on
hadron matrix elements of various quark and gluon operators.

The inclusive decay rate can be obtained using an operator product
expansion (OPE) to write the square of the decay amplitude as an
expansion in a series of local operators. The matrix elements of this
series between hadron states then gives the decay rate as an expansion
in $1/m_b$. The Wilson coefficients are functions of kinematic
variables $q^2$ and $v\cdot q$. The OPE can be justified for values of
$v\cdot q$ which are far from the physical region; the utility of the
expansion comes from relating a contour integral in the unphysical
region of the complex $v \cdot q$ plane to an integral over the
physical region~\cite{chay}. Unfortunately, the contour integral
always has a segment close to the physical region, where the OPE may
not be valid. Nevertheless, it is expected that the OPE computation of
the differential distributions (including non-perturbative
corrections) is valid, provided the results are smeared over an energy
of order $\Lambda_{\rm QCD}$. The idea that the parton calculation
agrees with the full QCD answer for {\it smeared} inclusive
distributions is known as local duality.

It is important to test the assumptions used in the OPE based
calculation of inclusive decay rates. One way to do this is to compare
the inclusive decay rate to the exclusive decay rate summed over all
allowed channels; they must be equal by definition. This equality will
arise nontrivially, because the OPE gives the inclusive rate in terms
of quark masses, rather than hadron masses.  Higher dimension
operators in the OPE must enter in precisely the correct way to
compensate for the mismatch. A demonstration that this occurs
alleviates many of the concerns about the validity of local duality in
QCD.

In this paper, we demonstrate by explicit computation that the
inclusive decay rates computed by an OPE or by summing over exclusive
rates are equal for hadrons containing a heavy quark in the
Shifman-Voloshin (SV) limit~\cite{SV} $m_b,m_c \gg \delta m=m_b-m_c
\gg \Lambda_{\rm QCD}$. The equality will be shown to hold to two
orders in the $1/m$ expansion, and to first order in $\alpha_s$, for
which explicit calculations exist in HQET.  In the SV limit, there are
$1/m$ corrections to the inclusive and exclusive decay rates. We
discuss the origin of these $1/m$ corrections, and show how they match
between the exclusive and inclusive decays. The differential decay
distributions for the inclusive and exclusive decay rates are also
shown to be equal, provided they are smeared over a region of size
$\Lambda_{\rm QCD}$. This is the expected size of the minimum smearing
region required before local duality holds in QCD.  We discuss the
variables that should be used when comparing inclusive and exclusive
decay distributions.  Finally, by studying the $1/m^2$ corrections, we
obtain inequalities on the slope of the Isgur-Wise function.
We also comment on hadronic matrix element inequalities obtained
earlier in the literature. The $\alpha_s/m^2$ corrections to the decay
width are used to obtain the $\alpha_s$ corrections to Bjorken's bound
on the slope of the Isgur-Wise function at zero recoil.

The kinematics of semileptonic $B$ decay are reviewed in
Sec.~\ref{sec:kin}.  The inclusive decay rate for $b\rightarrow c$ is
computed in Sec.~\ref{sec:inc}; the exclusive decays $B\rightarrow D$
and $B\rightarrow D^*$ are computed in Sec.~\ref{sec:exc}, and are
shown to agree to two orders in the $1/m$ expansion in the SV limit.
The electron spectrum and differential decay distributions for the
inclusive and exclusive decays are shown to agree to two orders in
$1/m$ in the SV limit in Sec.~\ref{sec:diff}.  The $\alpha_s$
corrections to the decay width and decay tensors are studied in
Sec.~\ref{sec:alphas}.  The $\alpha_s$ correction to Bjorken's bound
is derived in Sec.~\ref{sec:1/m2}.  We aslo study inequalities on
hadronic matrix elements in this section.  The results of earlier
sections are extended to $\Lambda_b$ decays in Sec.~\ref{sec:lambda}.

\section{Kinematics}\label{sec:kin}

In this section, we review some well-known results for the kinematics
of three-body decays which will be used extensively in the rest of
this paper.  We will consider the decay of a hadron $H_b$ containing a
$b$-quark. The formul\ae\ can be applied to the decay of a $b$-quark
by replacing the hadron mass $M_{H_b}$ by the quark mass, $m_b$.

Consider the decay $H_b \rightarrow X e^- \nu_e$, where $X$ is some
hadronic state with invariant mass $M_X$.\footnote{$X$ can be a
multi-particle state. The electron and neutrino masses will be
neglected for simplicity.}  The momenta of $H_b$, $X$, $e$ and $\nu_e$
will be denoted by $p_{H_b}$, $p_X$, $p_e$ and $p_\nu$ respectively,
and $q=p_{H_b}-p_X=p_e+p_\nu$ is the momentum transferred from the
hadron system to the leptons by the virtual $W$ boson. The velocity
four-vector $v$ is defined by $p_{H_b}=M_{H_b} v$, so that
$v=(1,0,0,0)$ defines the rest frame of the decay hadron. The hadronic
matrix elements for $H_b\rightarrow X e^- \nu_e$ can only depend on
the variables $q^2$, $q^0=q\cdot v$, and the masses $M_{H_b}$ and
$M_X$. A quantity of experimental interest is the electron energy
$E_e$ in the rest frame of the decaying hadron.

\begin{figure}
\moveright1.4in\hbox{\epsffile{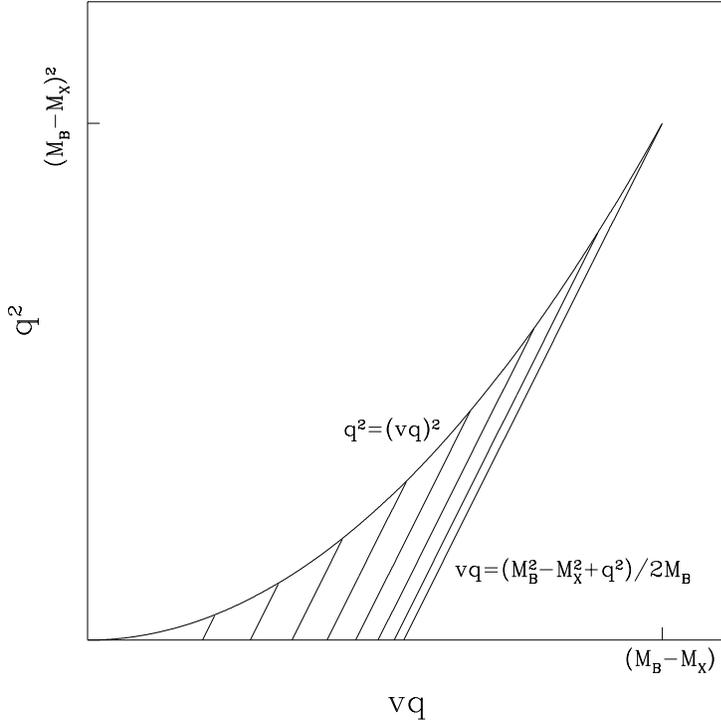}}
\caption{Plot of $q^2$ against $q\cdot v$. The slanted lines are the
allowed curves for different (equally spaced) values of $M_X$. The
upper edge of the allowed region is the zero recoil point for
different hadronic states $X$. The right-hand edge of the allowed
region is from the lightest allowed state $X$.}
\label{fig:q2qv}
\end{figure}

The momentum transfer to the hadronic system varies between $0 \le q^2
\le q^2_{\rm max}$, where
\begin{equation}\label{q2max}
q^2_{\rm max} = \left( M_{H_b} - M_X \right)^2.
\end{equation}
At $q^2=q^2_{\rm max}$ the final state hadronic system $X$ is at rest
in the rest frame of the decaying ${H_b}$ hadron. This kinematic point
is known as the zero-recoil point. The mass-shell condition $p_X^2 =
\left(p_{H_b}-q\right)^2 = M_X^2$ relates $q^2$ and $q\cdot v$,
\begin{equation}\label{q2qv}
q\cdot v = {M_{H_b}^2 - M_X^2 + q^2 \over 2 M_{H_b}}.
\end{equation}
As $q^2$ ranges from 0 to $q^2_{\rm max}$ for a fixed value of $M_X$,
$q\cdot v$ ranges from $\left(M_{H_b}^2 - M_X^2\right)/2 M_{H_b}$ to
$M_{H_b}-M_X$. The relation between $q^2 $ and $q\cdot v$ is plotted
in Fig.~\ref{fig:q2qv} for different values of the final state
hadronic mass $M_X$. The boundaries of the allowed region are
(including all allowed values of $M_X$)
\begin{mathletters}
\begin{eqnarray}
&&q^2 = \left(q\cdot v\right)^2,\\ \noalign{\smallskip} &&q^2=0,\ 0\le
q\cdot v \le {M^2_{H_b}-M_{\rm min}^2 \over 2 M_{H_b}} \\
\noalign{\smallskip} &&q\cdot v = {M_{H_b}^2 - M_{\rm min}^2 + q^2
\over 2 M_{H_b}}, \ 0 \le q^2 \le \left(M_{H_b}- M_{\rm min}\right)^2.
\end{eqnarray}
\end{mathletters}
The upper edge of the allowed region, $q^2 = (v\cdot q)^2$,
corresponds to the zero-recoil point for different hadronic
states. The minimum hadronic mass $M_{\rm min}$ is the pion mass for
$b\rightarrow u$ decays, and is the $D$ meson mass for $b\rightarrow
c$ decays.

\begin{figure}
\moveright1.4in\hbox{\epsffile{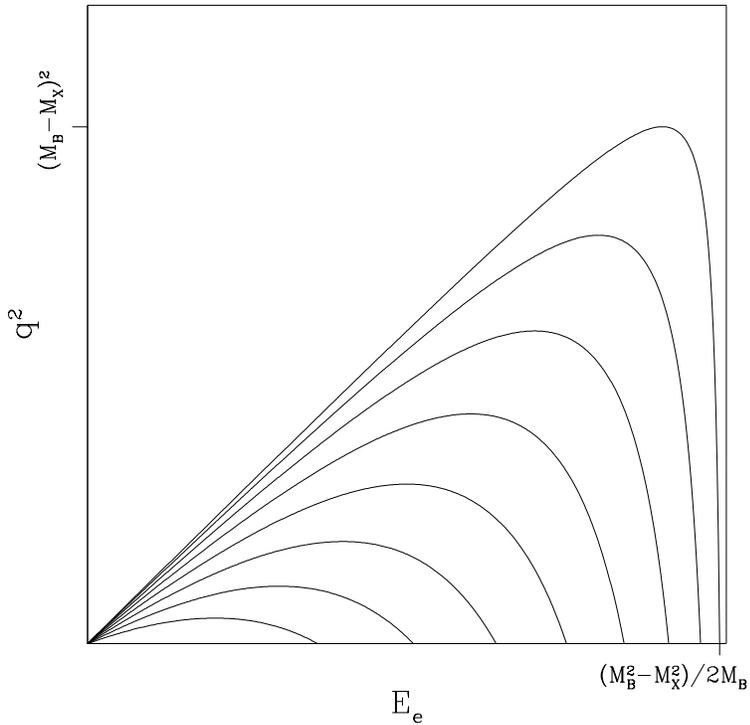}}
\caption{Plot of $q^2$ against the electron energy $E_e$. The allowed
region for different (equally spaced) values of the final state mass
$M_X$ is the interior of one of the curves. The total allowed region
is the interior of the largest curve, which corresponds to the lowest
allowed mass $M_X$. The zero recoil point for a given $M_X$ is at the
maximum allowed value of $q^2$.}
\label{fig:q2ee}
\end{figure}

In the $q^2$--$E_e$ plane, $E_e$ varies from 0 to
$(M_{H_b}^2-M_X^2)/2M_{H_b}$, and for a given value of $E_e$, the
$q^2$ range is
\begin{equation}
0 \le q^2 \le 2 E_e M_{H_b} - {2 E_e\,M_X^2 \over M_{H_b}- 2 E_e}.
\end{equation}
The maximum allowed value for $q^2$, Eq.~(\ref{q2max}), corresponds to
an electron energy of $(M_{H_b}-M_X)/2$. The allowed region in the
$q^2$--$E_e$ plane is plotted in Fig.~\ref{fig:q2ee}. For a given
hadronic mass $M_X$, the allowed region is the interior of one of the
curves.

In the SV limit, $M_{H_b},M_X\gg M_{H_b}-M_X \gg \Lambda_{\rm QCD}$,
the kinematically allowed regions for the variables for a fixed value
of $M_X$ are:
\begin{eqnarray}
&&0\le q^2 \le \left(\delta M_X\right)^2,\\ \noalign{\medskip}
&&\delta M_X-{\left(\delta M_X\right)^2 \over 2M_{H_b}} \le q\cdot v
\le \delta M_X,\\
&& 0 \le E_e \le \delta M_X-{\left(\delta M_X\right)^2 \over 2M_{H_b}}
\approx \delta M_X,
\end{eqnarray}
where $\delta M_X \equiv M_{H_b}-M_X$. The phase space volume for
three-body decay, which is proportional to $\int dq^2 dE_e$, is of
order $\left(\delta M_X\right)^3$ in the SV limit.

\section{Inclusive Decays}\label{sec:inc}

Semileptonic $b \to c$ decay is due to the weak hamiltonian density
\begin{eqnarray}\label{hw}
H_W &=& - V_{cb}\ {4 G_F\over \sqrt 2}\ \bar c \gamma^\mu P_L b\ \bar
e \gamma_\mu P_L \nu_e \\ &=& - V_{cb}\ {4 G_F\over \sqrt 2}\ J^\mu_h
J_{\ell \mu},\nonumber
\end{eqnarray}
where $P_L$ is the left handed projection operator $(1-\gamma_5)/2$,
and $J^\mu_h$ and $J^\mu_\ell$ are the hadronic and leptonic currents,
respectively.  The inclusive differential decay rate for a hadron
$H_b$ containing a $b$-quark to decay semileptonically,
$H_b\rightarrow X_{u,c}\, e\bar \nu_e$ is determined by the hadronic
tensor
\begin{eqnarray}\label{wmunu}
&&W^{\mu\nu} = \left(2\pi\right)^3\sum_X \delta^4\left(
p_{H_b}-q-p_X\right)\times \\ &&\langle H_b
(v,s)|J^{\mu\,\dagger}_h\left|X\right\rangle \left\langle
X\right|J^\nu_h\left|H_b(v,s)\right\rangle.\nonumber
\end{eqnarray}
The hadron state $\left|H_b(v,s)\right\rangle$ is normalized to $v^0$
instead of to the usual relativistic normalization of $2E$, as this is
more convenient for the heavy quark expansion.  $W^{\mu\nu}$ can be
expanded in terms of five form factors if one spin-averages over the
initial state,
\begin{eqnarray}\label{wi}
&& W^{\mu\nu}=-g^{\mu\nu} W_1 + v^{\mu}v^{\nu} W_2 - i
\epsilon^{\mu\nu\alpha\beta} v_\alpha q_\beta W_3 \\ && + q^\mu q^\nu
W_4 + \left(q^\mu v^\nu + q^\nu v^\mu\right) W_5. \nonumber
\end{eqnarray}
$W_1$ and $W_2$ have mass dimension $-1$, $W_3$ and $W_5$ have mass
dimension $-2$, and $W_4$ has mass dimension $-3$.  The form factors
are functions of the invariants $q^2$ and $q\cdot v$, and will also
depend on the initial hadron $H_b$ and the final quark mass $m_c$. The
spin averaged differential semileptonic decay rate is
\begin{eqnarray}\label{diffrate}
&&{d\Gamma\over dq^2\, dE_e\, dE_\nu} = {\left|V_{cb}\right|^2\,
G_F^2\over 2 \pi^3}\Bigl[ W_1\ q^2 \\ &&+ W_2 \left(2 E_e E_\nu -
{1\over2} q^2\right) + W_3\ q^2\left(E_e -
E_\nu\right)\Bigr],\nonumber
\end{eqnarray}
where $E_e$ and $E_\nu$ are the electron and neutrino energies in the
$H_b$ rest frame, $q^2$ is the invariant mass of the lepton pair, and
the kinematic variables are to be integrated over the region $q^2 \le
4 E_e E_\nu$.  The terms proportional to $q^\mu$ or $q^\nu$ in
Eq.~(\ref{wmunu}) do not contribute to the decay rate if one neglects
the electron mass.

The invariant tensors $W_i$ can be written as
\begin{equation}
W_i = -{1\over \pi} {\rm Im} T_i,
\end{equation}
where $T_i$ are defined via the hadronic matrix element of the
time-ordered product of the two currents,
\begin{eqnarray}
T^{\mu\nu} &=& -i\int d^4 x\ e^{-i q \cdot x} \langle H_b| T\left(
J^{\mu\,\dagger}_h\left(x\right) J^\nu_h\left(0\right)\right)
\left|{H_b}\right\rangle\nonumber\\ &=&-g^{\mu\nu} T_1 +
v^{\mu}v^{\nu} T_2 - i \epsilon^{\mu\nu\alpha\beta} v_\alpha q_\beta
T_3 + q^\mu q^\nu T_4\nonumber\\ && + \left(q^\mu v^\nu + q^\nu
v^\mu\right) T_5.\label{tdef}
\end{eqnarray}
The tensors $T_i$ have been computed to order
$1/m^2$~\cite{mannel,falk,MW,BKSV}. We will compute the inclusive and
exclusive rates not only for the case of physical interest, in which
the weak currents are left-handed, but also for more general
currents. For this reason, it is useful to break $T_i$ into the pieces
that arise from the time ordered product of two vector currents,
$T_i^{VV}$, two axial currents, $T_i^{AA}$, and one vector and axial
current $T_i^{AV}$, which have been computed
separately~\cite{BKSV}. The explicit expressions for the tensors are
given in Appendix~\ref{app:tensors} for completeness. The $1/m^2$
correction are written in terms of two hadronic matrix elements,
\begin{eqnarray}
G &=& Z_b \left\langle H_b(v) \right| \bar b_v
{g\sigma^{\mu\nu}G_{\mu\nu}\over4} b_v\left| H_b(v)
\right\rangle,\label{gdef}\\ \noalign{\smallskip} K &=& -\left\langle
H_b(v) \right| \bar b_v {\left(iD\right)^2\over2} b_v\left| H_b(v)
\right\rangle, \label{kdef}
\end{eqnarray}
which give the energy of the heavy quark in the hadron due to the
color magnetic moment interaction and kinetic energy,
respectively. $Z_b$ is a renormalization factor equal to unity at a
scale $\mu=m_b$.  In the $m_b\rightarrow\infty$ limit, $K$ and $G$ are
finite, and of order $\Lambda_{\rm QCD}^2$. The matrix elements $K$
and $G$ differ from the dimensionless matrix elements $K_b$ and $G_b$
of Ref.~\cite{MW} by a factor of $m_b^2$, $K = m_b^2 K_b$ and $G=m_b^2
G_b$.

The total inclusive decay rate for the case of $V-A$ decay is given by
using Eqs.~(\ref{teqns}), and has been obtained
previously~\cite{MW,BKSV,BSUV}, \widetext
\begin{eqnarray}
\Gamma^L\left(B\to X_c\right) &=& {G_F^2 m_b^5\over 192\pi^3}
\left|V_{cb}\right|^2\Biggl[ \left(1-8r + 8 r^3 - r^4-12r^2\log
r\right)+{G\over m_b^2} \left(3-8r+24r^2-24r^3+5r^4+12r^2\log
r\right)\nonumber\\ &&\qquad\qquad+{K\over
m_b^2}\left(-1+8r-8r^3+r^4+12r^2\log r\right)\Biggr],
\label{incform}
\end{eqnarray}
%
where $r=m_c^2/m_b^2$. Expanding the inclusive decay rate
in the SV limit for $b\rightarrow c$ decay gives
\begin{eqnarray}
&&\Gamma^L\left(B\to X_c\right) = {G_F^2\over 192\pi^3}
\left|V_{cb}\right|^2\Biggl[ {64\over 5} \left(\delta m\right)^5
{}-{96\over 5}{\left(\delta m\right)^6\over m_b}+ 64 {\left(\delta
m\right)^4 G\over m_b}\nonumber\\ &&+{32\over 35 m_b^2}\left(\delta
m\right)^5\left( 9 \left(\delta m\right)^2-154
G-14K+\ldots\right)\Biggr] + \ldots,\label{Lincexp}
\end{eqnarray}
where $\delta m = m_b-m_c$ is the difference of quark masses.  Terms
of order $1/m_b^3$ have been neglected. In addition, we have neglected
$1/m_b^2$ terms of order $\Lambda_{\rm QCD}^3/ \delta m\, m_b^2$. Such
terms can arise in the SV limit on expanding the $1/m_b^3$ corrections
to semileptonic decay, which have not been computed.

The leading order term in the decay rate has changed from ${G_F^2
m_b^5/192\pi^3}$ to ${G_F^2\left(\delta m\right)^5/15\pi^3}$. The
phase space volume is proportional to $\left(\delta m\right)^3$ in the
SV limit; the remaining $\left(\delta m\right)^2$ arises from factors
such as $q^2$ in Eq.~(\ref{diffrate}). In the limit
$m_b\rightarrow\infty$ with $\delta m$ fixed, the first term
approaches a constant. The second and third terms in
Eq.~(\ref{Lincexp}) are $1/m_b$ corrections to the decay rate, since
$\delta m$ and $G$ are both fixed in the SV limit as
$m_b\rightarrow\infty$. There are no $1/m_b$ corrections to the
tensors $T_i$, since the only possible operator that can occur in HQET
at order $1/m$, $\bar b (v \cdot D) b$ has vanishing matrix elements
by the equations of motion~\cite{chay}. Nevertheless, in the SV limit,
one still obtains a $1/m$ correction to the inclusive rate. The $G$
operator contribution to $T_i$ gives a term proportional to
$\left(\delta m\right)^4$ in the total rate, rather than $\left(\delta
m\right)^5$, which enhances the correction by one power of $m_b$ in
the SV limit. It is interesting that the $1/m_b$ correction depends
only on $G$, and not on $K$.  We will return to this point in the
section on exclusive decays.

One can also compute the total inclusive rate for a purely vector
current ``weak decay'' using $T_i^{VV}$. The resulting computation is
straightforward, and leads to the expression \widetext
\begin{eqnarray}
&&\Gamma^V\left(B\to X_c\right) = {G_F^2 m_b^5\over 96\pi^3}
\left|V_{cb}\right|^2\Biggl[ 1 - 2{\sqrt{r}} - 8r - 18{r^{{3/2}}} +
18{r^{{5/ 2}}} + 8{r^3} + 2{r^{{7/ 2}}} - {r^4} - 12\left({r^{{3/ 2}}}
+ {r^2}+r^{5/2}\right)\log r\nonumber\\ &&+ G\left( 3 +
{{26{\sqrt{r}}}\over 3} - 8r - 18{r^{{3/ 2}}} + 24{r^2} + 18{r^{{5/
2}}} - 24{r^3} - {{26{r^{{7/2}}}}\over 3} + 5{r^4} + \left(8{\sqrt{r}}
{}- 12{r^{{3/ 2}}} + 12{r^2} + 12{r^{{5/ 2}}}\right)\log r \right)
\nonumber\\ && + K\left( -1 + 2{\sqrt{r}} + 8r + 18{r^{{3/ 2}}} -
18{r^{{5/ 2}}} - 8{r^3} - 2{r^{{7/2}}} + {r^4} + 12\left({r^{{3/ 2}}}
+ {r^2} + {r^{{5/ 2}}}\right)\log r \right)\Biggr].
\end{eqnarray}
%
Expanding this in the SV limit gives
\begin{eqnarray}
&&\Gamma^V\left(B\to X_c\right) = {G_F^2\over 192\pi^3}
\left|V_{cb}\right|^2\Biggl[ {64\over 5} \left(\delta m\right)^5
{}-{96\over 5}{\left(\delta m\right)^6\over m_b}\nonumber\\ &&\qquad
+{32\over 35 m_b^2}\left(\delta m\right)^5\left( 12 \left(\delta
m\right)^2-70 G-14K+\ldots\right)\Biggr]\nonumber\\ &&\qquad +
\ldots.\label{Vincexp}
\end{eqnarray}
The structure is similar to the expansion of the left-handed current
in Eq.~(\ref{Lincexp}), except that the $1/m_b$ correction is
independent of any hadronic matrix elements. We will return to
Eqs.~(\ref{Lincexp}) and (\ref{Vincexp}) after we have computed the
exclusive decay rates.

\section{Exclusive Decays}\label{sec:exc}

In the SV limit for $B$ meson decay via $b\rightarrow c$ transitions,
the dominant hadronic final states are the $D$ and $D^*$. All other
states contribute to the decay rate only at order $1/m_b^2$ because of
heavy quark symmetry. At zero recoil, the leading order terms in the
$1/m$ expansion of the vector and axial currents are generators of the
heavy quark symmetry in the effective theory, so acting on the $B$
meson, they can only produce a linear combination of the $D$ and
$D^*$. The amplitude to produce excited states must either be of order
$1/m$, or vanish as $v\cdot v^\prime \rightarrow 1$. In either case,
the decay rate has at least a $1/m^2$ suppression factor.

An interesting example of this suppression is for the decays
$B\rightarrow D \pi e \nu$ and $B\rightarrow D^* \pi e \nu$. Chiral
perturbation theory can be used to compute the amplitude for these
decays when the pion momentum is small compared with the chiral
symmetry breaking scale $\Lambda_\chi$. One might expect that the
amplitude for producing the additional pion is proportional to ${\bf
p}_\pi/f_\pi$, and so is unsuppressed when ${\bf p}_\pi \sim
f_\pi$. However, an explicit computation~\cite{bbpi} shows that there
is a cancellation in the sum of the pion emission from the initial
meson, final meson, and vertex, and that the total amplitude at zero
recoil is proportional to $M_{B^*}-M_B$ or $M_{D^*}-M_D$, and is of
order $1/m$.

The $B\rightarrow D$ and $B \rightarrow D^*$ matrix elements of the
vector and axial currents can be parameterized in terms of six form
factors,
\begin{mathletters}\label{ffdefs}
\begin{eqnarray}
&&\left\langle D(p^\prime)\right| V^\mu \left|\bar B(p)\right\rangle =
\left(p+p^\prime\right)^\mu f_+ + \left(p-p^\prime\right)^\mu f_-,\\
\noalign{\smallskip} &&\left\langle D^*(p^\prime)\right| V^\mu
\left|\bar B(p)\right\rangle = i g \epsilon^{\mu\nu\alpha\beta}
\epsilon^*_{\nu}p^\prime_\alpha p_\beta,\\ \noalign{\smallskip}
&&\left\langle D^*(p^\prime)\right| A^\mu \left|\bar B(p)\right\rangle
= f_0 \epsilon^{*\mu} + \\ && \qquad\qquad\qquad \left[
\left(p+p^\prime\right)^\mu a_+ + \left(p-p^\prime\right)^\mu a_-
\right] p\cdot \epsilon^*,\nonumber
\end{eqnarray}
\end{mathletters}
where $f_{\pm}$, $f_0$, $a_{\pm}$ and $g$ are functions of $w=v\cdot
v^\prime$ (or equivalently, of $q^2$). The states in
Eq.~(\ref{ffdefs}) have the usual relativistic normalization of $2E$.

The contribution of the $D$ and $D^*$ final states to the hadronic
tensors $W_i$ can be computed by squaring the matrix elements
Eq.~(\ref{ffdefs}), and averaging over the $D^*$ polarizations. The
result for a left-handed current $L ={1\over 2}(V-A)$ is
\begin{eqnarray}\label{wd1}
&&8 M_B W_1 = 0,\\ &&8 M_B W_2 = 4 f_+^2\ M_B^2
\delta\left(\Delta_D\right),\label{wd2} \\ &&8 M_B W_3 = 0,
\label{wd3}\\ &&8 M_B W_4 = \left(f_+-f_-\right)^2\
\delta\left(\Delta_D\right),\label{wd4}\\ &&8 M_B W_5 = 2
f_+\left(f_--f_+\right)\ M_B \delta\left(\Delta_D\right),\label{wd5}
\end{eqnarray}
from $D$ final states, and

\widetext

\begin{eqnarray}\label{wds1}
&&8 M_B W_1 = \left[ g^2\left(p\cdot q^2 - M_B^2 q^2 \right) +
f_0^2\right] \delta\left(\Delta_{D^*}\right),\\ \noalign{\medskip} &&8
M_B W_2 = \left[-q^2 g^2 + {f_0^2\over M_{D^*}^2} + 4
a_+^2\left(-M_B^2 + { \left(M_B^2-p\cdot q\right)^2\over
M_{D^*}^2}\right) + f_0\ a_+\left( -4 + 4 {M_B^2\over M_{D^*}^2} - 4
{p\cdot q \over M_{D^*}^2}\right)\right] M_B^2\,
\delta\left(\Delta_{D^*}\right), \label{wds2}\\ \noalign{\medskip} &&8
M_B W_3 = 2\, g\, f_0\
M_B\,\delta\left(\Delta_{D^*}\right),\label{wds3} \\
\noalign{\medskip} &&8 M_B W_4 = \left[-g^2 M_B^2 + {f_0^2\over
M_{D^*}^2} + \left(a_+-a_-\right)^2\left( -M_B^2 +{\left(M_B^2-p\cdot
q\right)^2\over M_{D^*}^2}\right)+ 2 f_0
\left(a_+-a_-\right){M_B^2-p\cdot q\over M_{D^*}^2}\right]
\delta\left(\Delta_{D^*}\right) ,\label{wds4}\\ \noalign{\medskip} &&8
M_B W_5 =\left[p\cdot q\ g^2 - {f_0^2\over M_{D^*}^2} + 2 a_+^2
\left(M_B^2 - { \left(M_B^2-p\cdot q\right)^2\over M_{D^*}^2}\right) +
f_0\ a_+\left( 1-3{\left(M_B^2-p\cdot q\right)\over
M_{D^*}^2}\right)\right. + \nonumber\\ &&\qquad\qquad\qquad\left. f_0\
a_- \left({\left(M_B^2-p\cdot q\right)\over M_{D^*}^2}-1\right) + 2
a_+\ a_- \left(-M_B^2 + { \left(M_B^2-p\cdot q\right)^2\over
M_{D^*}^2}\right) \right] M_B\,
\delta\left(\Delta_{D^*}\right)\label{wds5}
\end{eqnarray}
%
from $D^*$ final states. $\Delta_{D,D^*}$ are defined by
\begin{eqnarray}\label{ddddsdef}
\Delta_D &=& \left(p_B-q\right)^2 - M_D^2,\\ \Delta_{D^*} &=&
\left(p_B-q\right)^2 - M_{D^*}^2.
\end{eqnarray}
Note that all the form factors in Eqs.~(\ref{wd1})--(\ref{wds5})
depend implicitly on $q^2$. The $W$'s for a vector current or an axial
current can be obtained trivially from Eqs.~(\ref{wd1})--(\ref{wds5})
by setting either the axial or vector form factors to zero, and
multiplying by four, since $L^2=(V-A)^2/4$.

The decay rate for $B\rightarrow D$ and $B\rightarrow D^*$ is obtained
by integrating $W_i$ (or equivalently, the square of the matrix
elements in Eq.~(\ref{ffdefs})) over the allowed kinematic region
using Eq.~(\ref{diffrate}).
\begin{eqnarray}
w-1=v\cdot v^\prime - 1 &=& {(M_B-M_D)^2 - q^2 \over 2 M_B
M_D},\nonumber\\ &=& {q^2_{\rm max} - q^2\over 2 M_B M_D},
\end{eqnarray}
so that in the SV limit $w-1$ is of order $1/M^2$ over the entire
kinematic region for the decay process. Thus we can expand the form
factors in a series in $w-1$,
\begin{equation}
f\left(w\right) = f\left(1\right) + \left(w-1\right){df\over
dw}\left(1\right) + \ldots \label{ffexpand}
\end{equation}
where $f$ denotes any of the six form factors in Eq.~(\ref{ffdefs}).

For left-handed currents, the decay rate for $B\rightarrow D$ is
\widetext
\begin{eqnarray}
&&\Gamma^L\left(B\to D\right)= {G_F^2\over 192\pi^3}
\left|V_{cb}\right|^2\Biggl[ {16\over 5} \left(\delta M\right)^5
f_+^2-{24\over 5}{\left(\delta M\right)^6\over M_B} f_+^2 +{8\over 35
M_B^2} \left(\delta M\right)^7 f_+\left( 10 {df_+\over dw} +9
f_+\right)+ \ldots \Biggr],\label{Ldexp}
\end{eqnarray}
and for $B\rightarrow D^*$ is
\begin{eqnarray}
&&\Gamma^L\left(B\to D^*\right) = {G_F^2\over 192\pi^3}
\left|V_{cb}\right|^2\Biggl[ {12\over 5} \left(\delta M^*\right)^5
{f_0^2\over M_B^2}+{2\over 5}{\left(\delta M^*\right)^6\over
M_B}f_0\left(8 a_+ - {f_0\over M_B^2}\right)\nonumber\\ &&+{2\over 35
M_B^2}\left(\delta M^*\right)^7\Bigl(40 M_B^2 a_+^2 - 8 a_+f_0 + 22
{df_0 \over dw} {f_0 \over M_B^2} + 7 {f_0^2\over M_B^2} + 8 M_B^2
g^2\Bigr)+\ldots\Biggr],\label{Ldsexp}
\end{eqnarray}
%
where
\begin{eqnarray}
\delta M &=& M_B - M_D, \label{deltam}\\ \delta M^* &=& M_B - M_{D^*},
\label{deltams}
\end{eqnarray}
are hadron mass differences, and all form factors are evaluated at
$w=1$. The neglected terms are of order $1/M_B^3$. Equations
(\ref{Ldexp}) and (\ref{Ldsexp}) are important results from this
section, and will be used repeatedly in the rest of this paper. The
decay rates for a purely vector current weak decay are obtained by
multiplying Eqs.~(\ref{Ldexp}-\ref{Ldsexp}) by four, and setting the
axial form factors to zero, \widetext
\begin{eqnarray}
&&\Gamma^V\left(B\to D\right)= {G_F^2\over 192\pi^3}
\left|V_{cb}\right|^2\Biggl[ {64\over 5} \left(\delta M\right)^5 f_+^2
{}-{96\over 5}{\left(\delta M\right)^6 \over M_B} f_+^2 +{32\over 35
M_B^2} \left(\delta M\right)^7 f_+\left( 10 {df_+\over dw} +9
f_+\right) + \ldots \Biggr],\label{Vdexp}
\end{eqnarray}
\begin{equation}
\Gamma^V\left(B\to D^*\right) = {G_F^2\over 192\pi^3}
\left|V_{cb}\right|^2\Biggl[ {64\over 35}\left(\delta M^*\right)^7
g^2+\ldots\Biggr].\label{Vdsexp}
\end{equation}

The hadron masses can be computed in terms of quark masses in HQET to
order $1/m^2$,
\begin{eqnarray}
M_B &=& m_b + \bar \Lambda + {K\over m_b} + {G\over m_b} +
{K^\prime\over m_b^2} +{G^\prime\over m_b^2} + \ldots,\label{Bmass}\\
M_D &=& m_c + \bar \Lambda + {K\over m_c} + {G\over m_c}+
{K^\prime\over m_c^2}+ {G^\prime \over m_c^2}+ \ldots,\label{Dmass}\\
M_{B^*} &=& m_b + \bar \Lambda + {K\over m_b} - {G\over 3m_b} +
{K^\prime\over m_b^2} -{G^\prime\over 3m_b^2}+ \ldots,\label{Bsmass}\\
M_{D^*} &=& m_c + \bar \Lambda + {K\over m_c} - {G\over 3m_c} +
{K^\prime\over m_c^2} - {G^\prime\over 3m_c^2}+ \ldots\label{Dsmass},
\end{eqnarray}
where $K$ and $G$ are defined in
Eqs.~(\ref{gdef}-\ref{kdef}). $K^\prime$ and $G^\prime$ are the
spin-independent and spin-dependent splittings at order $1/m^3$. The
hadron mass difference $\delta M$ is\footnote{G for $B$ and $D$ mesons
differ by anomalous scaling between $\mu=m_b$ and $\mu=m_c$. This
affects $\delta M$ at order $\alpha_s G \delta m/m_b^2$, and is
neglected here.}
\begin{eqnarray}
\delta M &=& M_B - M_D = m_b-m_c + \left(K+G\right)\left({1\over m_b}
{}-{1\over m_c}\right)\nonumber\\ &=& \delta m \left[ 1 - {K+G\over
m_b^2}\right] + {\cal O}\left({1\over m_b^3}\right),\label{dM}
\end{eqnarray}
in the SV limit. Note that the hadron mass difference $\delta M$ is
equal to the quark mass difference $\delta m$ to order $1/m^2$.
Similarly,
\begin{eqnarray}
&&\delta M^* = M_B - M_{D^*} \nonumber\\ &&= m_b-m_c +
\left(K+G\right)\left({1\over m_b} -{1\over m_c}\right) + {4\over3}
{G\over m_c}\nonumber\\ &&= \delta m \left[ 1 + {4G\over3 m_b\,\delta
m } - {K+G\over m_b^2}+{4 G^\prime\over 3 m_b^2 \delta m}\right] +
{\cal O}\left({1\over m_b^3}\right),\label{dMs}
\end{eqnarray}
in the SV limit. There is a $1/m$ term in the relation between hadron
and quark mass differences for the $D^*$, which arises because of the
$D^*-D$ mass difference, and is proportional to the matrix element
$G$. This $1/m$ term is required to reproduce the correct $1/m$
correction to the exclusive rate in Eq.~(\ref{Lincexp}), and explains
why the $1/m$ correction in that formula depended only on the matrix
element $G$, and not on $K$.

To compare with the exclusive rate, we also need the form factors at
zero recoil in an expansion in $1/m_b$~\cite{Luke},
\begin{mathletters}\label{ffexp}
\begin{eqnarray}
f_+(1) &=& {M_B+M_D\over 2\sqrt{M_B M_D}}\left[1 + f_+^{(2)}{1\over
m_b^2}+\ldots\right] ,\\ \noalign{\smallskip} f_0(1) &=& 2\sqrt{M_B
M_{D^*}}\left[1 + f_0^{(2)}{1\over m_b^2}+\ldots\right] ,\\
\noalign{\smallskip} a_+(1) &=& -{1\over 2}\sqrt{{1\over M_B
M_{D^*}}}\left[1 +a_+^{(1)} {1\over m_b} + \ldots \right] ,\\
\noalign{\smallskip} g(1) &=& \sqrt{{1\over M_B M_{D^*}}}\left[1 +
\ldots \right] .
\end{eqnarray}
\end{mathletters}
The slopes of $f_0$ and $f_+$ at zero recoil ${df_0/ dw}(1),{df_+/
dw}(1) $ are simply related to the slope $-\rho^2$ of the Isgur Wise
function $\xi(w) = \xi(1) - \rho^2 (w -1) + {\cal O}(w-1)^2$ by
\begin{eqnarray}\label{slopetorho}
{df_0\over dw}(1)&=& \sqrt{M_B M_{D^*}} (1 - 2 \rho^2) + {\cal
O}(\delta m), \nonumber\\ {df_+\over dw}(1)&=& -\rho^2 + {\cal
O}(\delta m).
\end{eqnarray}
Substituting Eqs.~(\ref{dM}), (\ref{dMs}), (\ref{ffexp}) and
(\ref{slopetorho}) into Eqs.~(\ref{Ldexp}) and ~(\ref{Ldsexp}), and
adding, gives the total rate \widetext
\begin{eqnarray}
&&\Gamma^L\left(B\to D+ D^*\right) = {G_F^2\over 192\pi^3}
\left|V_{cb}\right|^2\Biggl[ {64\over 5} \left(\delta M\right)^5
{}-{96\over 5}{\left(\delta M\right)^6\over M_B} +64 { G\left(\delta
M\right)^4\over M_B} \nonumber\\ && +{8\over 35 M_B^2}\left(\delta
M\right)^5\Bigl( 44 \left(\delta M\right)^2 -32 \rho^2 \left(\delta
M\right)^2 -14 a_+^{(1)} \delta M + 28 f_+^{(2)} + 84 f_0^{(2)} - 224
G + 280 {G^\prime\over \delta M} + 280 {G\bar \Lambda\over \delta M}
+{2240\over3} {G^2\over \left(\delta M\right)^2}\Bigr)\Biggr]
\nonumber \\ &&\qquad\qquad\qquad+ \ldots.\label{Lexc}
\end{eqnarray}
Similarly, the total decay rate for vector current weak decay is given
by
\begin{eqnarray}
&&\Gamma^V\left(B\to D+ D^*\right) = {G_F^2\over 192\pi^3}
\left|V_{cb}\right|^2\Biggl[ {64\over 5} \left(\delta M\right)^5
{}-{96\over 5}{\left(\delta M\right)^6\over M_B}+{32\over 35
M_B^2}\left(\delta M\right)^5\left( {29\over2} \left(\delta M\right)^2
{}- 10 \rho^2 \left(\delta M\right)^2 + 28 f_+^{(2)} \right)\Biggr] +
\ldots.\label{Vexc}
\end{eqnarray}

It is useful to rewrite the inclusive decay rates Eqs.~(\ref{Lincexp})
and (\ref{Vincexp}) using the hadron mass difference $\delta M$,
rather than the quark mass difference $\delta m$. The expressions are
\begin{eqnarray}
&&\Gamma^L\left(B\to X_c\right) = {G_F^2\over 192\pi^3}
\left|V_{cb}\right|^2\Biggl[ {64\over 5} \left(\delta M\right)^5
{}-{96\over 5}{\left(\delta M\right)^6\over M_B}+ 64 {\left(\delta
M\right)^4 G\over M_B}\nonumber\\ &&+{32\over 35 M_B^2}\left(\delta
M\right)^5\left( 9 \left(\delta M\right)^2-21 \bar \Lambda
\left(\delta M\right) - 84 G +56 K + {70 G\bar\Lambda\over\delta
M}+\ldots\right)\Biggr] + \ldots,\label{Linc2}
\end{eqnarray}
\begin{eqnarray}
&&\Gamma^V\left(B\to X_c\right) = {G_F^2\over 192\pi^3}
\left|V_{cb}\right|^2\Biggl[ {64\over 5} \left(\delta M\right)^5
{}-{96\over 5}{\left(\delta M\right)^6\over M_B}+{32\over 35
M_B^2}\left(\delta M\right)^5\left( 12 \left(\delta M\right)^2-21
\bar\Lambda \delta M+56 K+\ldots\right)\Biggr] + \ldots, \label{Vinc2}
\end{eqnarray}

%
The exclusive decay rates Eqs.~(\ref{Lexc}) and
(\ref{Vexc}) agree with the inclusive decay rates Eqs.~(\ref{Linc2})
and (\ref{Vinc2}) to two orders in the $1/m$ expansion. The leading
order term in the hadronic decay rate is proportional to $\left(\delta
M\right)^5$ or $\left(\delta M^*\right)^5$. This functional dependence
is essential for the exclusive and inclusive decay rates to match.
The parameter $\bar\Lambda$ cancels in $\delta M$ and $\delta M^*$, so
$\bar\Lambda$ does not enter the relation between the inclusive and
exclusive rates. The $1/m$ corrections to the rates also match between
the inclusive and exclusive decays. The $1/m$ matching is non-trivial
in the case of $V-A$ decay, where it depends on the matrix element
$G$. The $G$ dependence arises from the $1/m$ term in $\delta M^*$ in
Eq.~(\ref{Ldsexp}), and implicitly through the $M_{D^*}$ dependence in
the form factors Eq.~(\ref{ffexp}). Agreement between the inclusive
and exclusive decays at order $1/m$ strongly suggests that agreement
will persist to all orders in the $1/m$ expansion including
non-perturbative corrections.

At order $1/m^2$, the analysis is more complicated because excited
states contribute to the exclusive decays. Nevertheless, one can still
obtain a useful inequality, since the $B$ partial decay width to
excited states is positive, so that
\begin{equation}
\Gamma^L\left(B\to X_c\right) \ge \Gamma^L\left(B\to D+D^*\right).
\end{equation}
This inequality implies that \widetext
\begin{eqnarray}
{32\over 35}\left( 9 \left(\delta M\right)^2-21 \bar \Lambda
\left(\delta M\right) - 84 G +56 K + {70 G\bar\Lambda\over\delta
M}\right) &\ge& {8\over 35}\Bigl( 44 \left(\delta M\right)^2 -32
\rho^2 \left(\delta M\right)^2 -14 a_+^{(1)} \delta M + 28 f_+^{(2)} +
84 f_0^{(2)}\nonumber\\ && - 224 G + 280 {G^\prime\over \delta M} +
280 {G\bar \Lambda\over \delta M} +{2240\over3} {G^2\over \left(\delta
M\right)^2}\Bigr).
\end{eqnarray}
%
The parameters $G$, $K$, $\bar\Lambda$, $a_+^{(1)}$ and
$f_+^{(2)}$ are all of order $\Lambda_{\rm QCD}^2$. Since $\delta M\gg
\Lambda_{\rm QCD}$, one can neglect those terms to obtain the Bjorken
inequality on the slope of the meson Isgur-Wise function~\cite{bj}
\begin{equation}
\rho^2 > {1\over4}.
\end{equation}
We will obtain the $\alpha_s$ corrections to the Bjorken inequality
using the $\alpha_s$ corrections to the inclusive and exclusive decay
rates in Sec.~\ref{sec:alphas}.

\section{Differential Distributions}\label{sec:diff}

In this section we will show that all the differential decay
distributions for semileptonic $B\to X_c$ and $B\to D+D^*$ decay agree
to two orders in the $1/m_b$ expansion, provided they are smeared over
a region of size $\Lambda_{\rm QCD}$.  We will first demonstrate this
for the electron spectrum, and then generalize the result to the
tensors $T_i$. The equality of the tensors $T_i$ means that all
moments of the inclusive and exclusive distributions agree to order
$1/m^2$. The comparison of $T_i$ for inclusive and exclusive decays
sheds some light on how local duality arises in QCD. It also specifies
which variables should be used when comparing the inclusive and
exclusive decay distributions.  We will only give the equations for
the physically interesting case of $V-A$ interactions; it is
straightforward to obtain the corresponding results for a vector
current decay.

\subsection{Electron Spectrum}\label{ssec:espectrum}

The electron spectrum in semileptonic $b$ decay can be computed using
the tensors $T_i$, and Eq.~(\ref{diffrate}). Expanding the known
results~\cite{MW,BKSV,BSUV} in the SV limit gives
\begin{eqnarray}
{ d\Gamma \over d E_e} = {2 G_F^2 \over \pi^3} \left|V_{cb}\right|^2
      E_e^2 \left(E_e - \delta M\right) \Biggl[ \left(E_e - \delta
      M\right) + \nonumber \\ {1\over M_B}\left( 2 \left(\delta
      M\right)^2 - 5 E_e \delta M + 4 E_e^2 - 2 G \right)
      \Biggr].\label{einc}
\end{eqnarray}
The endpoint of the inclusive electron spectrum is obtained using
quark masses,
\begin{equation}
E_{{\rm end},q} = {m_b^2-m_c^2 \over 2 m_b}.
\end{equation}

The electron spectrum from $B \to D$ decay is
\begin{eqnarray}
{ d\Gamma \over d E_e} = { G_F^2 \over 2 \pi^3} \left|V_{cb}\right|^2
          E_e^2 \left(E_e - \delta M\right) \Biggl[ \left(E_e - \delta
          M\right) + \nonumber \\ {1\over M_B}\left( \left(\delta
          M\right)^2 -2 E_e \delta M + 2 E_e^2 \right)
          \Biggr],\label{eD}
\end{eqnarray}
and from $B \to D^*$ decay is
\begin{eqnarray}
{ d\Gamma \over d E_e} = { G_F^2 \over 2 \pi^3} \left|V_{cb}\right|^2
           E_e^2 \left(E_e - \delta M^*\right) \Biggl[ 3\left(E_e -
           \delta M^*\right) + \nonumber \\ {1\over M_B}\left( 7
           \left(\delta M^*\right)^2 - 18 E_e \delta M^* + 14
           E_e^2\right) \Biggr].\label{eD*}
\end{eqnarray}
The endpoints of the electron spectrum for $B \to D$ and $B \to D^*$
 semileptonic decay are
\begin{equation}
E_{{\rm end},D} = {M_B^2-M_D^2\over 2 M_B},\ \ E_{{\rm end},D^*} =
{M_B^2-M_{D^*}^2 \over 2 M_B},
\end{equation}
respectively.

The electron spectrum from the inclusive decays, Eq.~(\ref{einc})
agrees with the sum of the electron spectrum from $B\rightarrow D$ and
$B\rightarrow D^*$ from Eq.~(\ref{eD}) and Eq.~(\ref{eD*}). The $G$
term arises when $\delta M^*$ is rewritten in terms of $\delta M$
using Eq.~(\ref{dMs}). It is important to note that the spectra agree
when written in terms of a measurable quantity, such as the electron
energy $E_e$. The spectra would not agree at order $1/m$ if written in
terms of dimensionless variables, such as $y_{\rm incl}=2E_e/m_b$ and
$y_{\rm excl}= 2E_e/m_B$, because there is a $1/m$ correction
proportional to $\bar\Lambda$ in $m_b/M_B$.

The endpoint of the inclusive electron spectrum agrees with the
endpoint of the $B\rightarrow D$ spectrum to two orders in $1/m$,
\begin{equation}
E_{{\rm end},D} =E_{{\rm end},q}\left[1 + {\cal O}\left({1\over
m_b^2}\right)\right],
\end{equation}
but the endpoint of the $B\rightarrow D^*$ spectrum differs at order
$1/m$,
\begin{equation}
E_{{\rm end},D^*} = E_{{\rm end},q}\left[1 + {4G\over3m_b \delta m}+
{\cal O}\left({1\over M_B^2}\right)\right]
\end{equation}
However, the leading term in the $B\rightarrow D^*$ spectrum vanishes
quadratically at its endpoint, and the first correction vanishes
linearly.  Thus, the difference between the quark and hadron endpoints
does not enter the expression for the electron spectrum until the
$1/m^2$ correction, and the two spectra agree point-by-point to order
$1/m$, without any smearing.

\subsection{The tensors $T_i$}\label{ssec:ti}

We will now compare the differential decay distributions for $B\to
X_c$ and $B\to D+D^*$ decays. All the information in the differential
decay distributions for arbitrary decays is contained in the hadronic
tensor $W_{\mu\nu}$ for $VV$, $AV$ and $AA$ currents. We will compare
these for inclusive and exclusive decays, and show that they agree to
two orders in $1/m$. The tensor $W_{\mu\nu}$ can be expanded into
Lorentz invariant functions $W_i$ defined in Eq.~(\ref{wi}).  Since
$q$ is of order $\delta m$ in the SV limit, $W_1$ and $W_2$ must match
to two orders in $1/m$; $W_3$ and $W_4$ must match to first order in
$1/m$; $W_4$ does not have to match. The inclusive and exclusive decay
tensors only match when suitably smeared. This is obvious, because
$W_i$ for the exclusive decays contains only $\delta$-functions,
whereas $W_i$ for the inclusive decay contains $\delta$-functions, as
well as their first and second derivatives. The gradients of $\delta$
functions are an approximation to a $\delta$ function of shifted
argument, $\delta(x+a) =\delta(x) + a \delta^\prime(x)+\ldots$. The
presence of singular functions such as the delta function and its
derivatives requires that the inclusive and exclusive decay tensors be
smeared, so that the correction terms $\delta^\prime(x)$ are
``smaller'' than the leading order term $\delta(x)$.

In comparing the inclusive with exclusive decays, it is convenient to
use the spectral representation to define the time ordered product
$T_{\mu\nu}$ for exclusive decays, and to compute the tensors $T_i$,
rather than their imaginary parts $W_i$. The tensors $T_i$ can be
obtained trivially from $W_i$ by the replacement
\begin{eqnarray}\label{wtconv}
\delta\left(\Delta_D\right) \rightarrow {1\over \Delta_D},\\
\delta\left(\Delta_{D^*}\right) \rightarrow {1\over \Delta_{D^*}},\\
\end{eqnarray}
in Eqs.~(\ref{wd1})--(\ref{wds5}), where $\Delta_{D,D^*}$ are defined
in Eq.~(\ref{ddddsdef}).

The tensors $T_i$ for inclusive decay are given in
Eqs.~(\ref{teqns}). We can expand them in the SV limit, treating
$\Delta_q = m_b^2-m_c^2-2 m_b q \cdot v + q^2$ as typically of order
$m_b \Lambda_{\rm QCD}$. The physical region corresponds to
$\Delta_q=0$. Assuming $\Delta_q \sim m_b\Lambda_{\rm QCD}$ means that
one is working in the complex $q\cdot v$ plane at least a distance
$\Lambda_{\rm QCD}$ away from the physical cut (See
Fig.~\ref{fig:qvcut}). This implies that one can only compute physical
quantities smeared over an energy range of order $\Lambda_{\rm
QCD}$~\cite{PQW}. Expanding Eqs.~(\ref{teqns}) gives
\begin{eqnarray}\label{qtensor}
T^{VV}_1 &=& {\delta m-q\cdot v\over\Delta_q} + {\cal O}\left(
{\Lambda\over m_b^2}\right),\\ T^{VV}_2 &=& 2{m_b\over\Delta_q} +
{\cal O}\left( {\Lambda\over m_b^2}\right),\\ T^{VV}_3 &=&0,\\
T^{VV}_4 &=& {\cal O}\left( {1\over m_b^3}\right),\\ T^{VV}_5 &=&
{}-{1\over\Delta_q}+ {\cal O}\left({1\over m_b^2}\right),\\ T^{AA}_1 &=&
2 {m_b\over\Delta_q} - \left({\delta m+q\cdot v\over\Delta_q}
+{16\over3\Delta_q^2} m_b G\right)\nonumber\\ && + {\cal O}\left(
{\Lambda\over m_b^2}\right),\\ T^{AA}_2 &=&
2{m_b\over\Delta_q}-{16\over3\Delta_q^2}m_b G + {\cal O}\left(
{\Lambda\over m_b^2}\right),\\ T^{AA}_3 &=&0,\\ T^{AA}_4 &=& {\cal
O}\left( {1\over m_b^3}\right),\\ T^{AA}_5 &=& -{1\over\Delta_q}+
{\cal O}\left({1\over m_b^2}\right),\\ T^{AV}_{i\not=3} &=& 0,\\
T^{AV}_3 &=& -{1\over\Delta_q} + {\cal O}\left( {1\over m_b^2}\right),
\end{eqnarray}
where $\Lambda\sim \Lambda_{\rm QCD}$ or $\sim\delta m$.  The
corresponding expansions of $T_i$ for $B\rightarrow D$ are
\begin{eqnarray}
T_1^{VV} &=& 0,\\ T_2^{VV} &=& {2M_B\over\Delta_D} + {\cal
O}\left({\Lambda\over M_B^2}\right),\\ T_3^{VV} &=& 0,\\ T_4^{VV} &=&
{\cal O}\left( {1\over \Lambda M_B^2}\right),\\ T_5^{VV} &=&
{}-{1\over\Delta_D}+{\cal O}\left( {1\over M_B^2}\right),\\ T_{i}^{AA}
&=& 0,\\ T_i^{AV} &=& 0,
\end{eqnarray}
and for $B\rightarrow D^*$ are
\begin{eqnarray}
T_1^{AA} &=& 2{M_B-\delta M^*\over\Delta_{D^*}} + {\cal
O}\left({\Lambda\over M_B^2}\right),\\ T_2^{AA} &=& {2M_B+q\cdot
v-\delta M^*\over \Delta_{D^*}} + {\cal O}\left({\Lambda\over
M_B^2}\right),\\ T_3^{AA} &=& 0 ,\\ T_4^{AA} &=& {\cal O}\left(
{1\over\Lambda M_B^2}\right),\\ T_5^{AA} &=& -{1\over \Delta_{D^*}}
+{\cal O}\left({1\over M_B^2}\right),\\ T_{i\not=3}^{AV} &=& 0,\\
T_3^{AV} &=& -{1\over\Delta_{D^*}} + {\cal O}\left({1\over
M_B^2}\right),
\end{eqnarray}
and $T_i^{VV}$ are all zero to this order. In deriving these
equations, we have assumed that $\Delta_D$ and $\Delta_{D^*}$ are of
order $M_B \Lambda_{\rm QCD}$.

Using the relation between the $D^*$ and $D$ masses Eq.~(\ref{dMs}),
one can write
\begin{equation}
\Delta_{D^*} = \Delta_D + {8\over3} G
\end{equation}
so that
\begin{equation}\label{Dexp}
{1\over\Delta_{D^*}} = {1\over\Delta_D} - {8\over3} {G\over
\Delta_D^2} +\ldots
\end{equation}
The expansion is allowed since we are assuming that $\Delta_D$ is of
order $M_B \Lambda_{\rm QCD}$.  We also need the relation
\begin{equation}\label{Dreln}
{ m_b\over \Delta_q} = {M_B \over \Delta_D}\left[1+{\cal
O}\left({1\over m_b^2}\right)\right].
\end{equation}

One can now compare $T_i$ for inclusive and exclusive decays. Most of
the $T_{1,2}$'s agree to order $1/m^2$ and $T_{3,5}$'s agree to order
$1/m$.  The only subtlety is that each of the tensors $T_1^{VV}$, and
$T_{1,2}^{AA}$ for inclusive decay is equal to the corresponding
tensor for exclusive decay plus the term
\begin{equation}
{\delta m - q\cdot v\over\Delta_q}.
\end{equation}
This term can be rewritten as
\begin{equation}
{1\over2m_b} + {\left(\delta m\right)^2-q^2\over2m_b\Delta_q}
\end{equation}
using the identity $\Delta_q = 2 m_b\left(\delta m - q\cdot
v\right)+q^2-\left(\delta m\right)^2$. The second term is of order
$1/m_b^2$, and is of the same order as terms we have neglected.  The
first term is of order $1/m$, and must be retained. {\it However, it
is analytic, and has no imaginary part.} Thus $W_{\mu\nu}$, which is
the imaginary part of $T_{\mu\nu}$ and describes the decay
distributions, agrees between the inclusive and exclusive decays at
first two orders in the $1/m$ expansion.

The above calculation shows that the first two terms of all the decay
distributions must match between the inclusive and exclusive
decays. It also shows that the distributions match when considered as
a function of $q\cdot v$ and $q^2$. They would not agree at order
$1/m$ if written in terms of $p\cdot q$ and $q^2$, since $m_b$ and
$M_B$ differ at order $1/m$. We also required $\Delta_{q,D}\sim M_B
\Lambda_{\rm QCD}$ and the expansion Eq.~(\ref{Dreln}) to obtain an
equality between the inclusive and exclusive decay distributions.
This means that one must smear over a region of order $\Lambda_{\rm
QCD}$ before duality between the parton and hadron calculations will
hold. In particular, the smearing region must be much larger than the
$D^*$--$D$ mass difference of 140~MeV, which is of order $\Lambda_{\rm
QCD}^2/m$.  Finally, note that $T^{\mu\nu}$ for inclusive and
exclusive decays can differ by polynomial terms with no imaginary
parts. This corresponds to having possible subtraction constants in
dispersion relations for $W_i$.

\section{$\alpha_s$ Corrections}\label{sec:alphas}

The order $\alpha_s\left(m_b\right)$ corrections to the total weak
decay rate for $b$ quarks has been computed~\cite{alphas}. The
$\alpha_s$ corrections that depend on $\alpha_s\left(m_b\right)$
should match between the inclusive and exclusive decay rates.  We will
see that this is indeed true, by explicit calculation. The agreement
between the inclusive and exclusive decays is very interesting,
because the first three terms have the form $\left(\delta m\right)^5$,
$\left(\delta m\right)^6$, and $\left(\delta m\right)^6 \log
\left(\delta m\right)$.  All three terms must match, since the excited
states first appear at order $\left(\delta m\right)^7$. We study the
differential decay distributions for inclusive and exclusive decays,
and show how they agree to order $\alpha_s$. We then use the
positivity of the partial width into excited states to obtain the
$\alpha_s$ correction to Bjorken's inequality.

\subsection{The Decay Width}

The total weak decay rate for $b$ quark decay can be written as
\begin{equation}\label{incalpha}
\Gamma = {G_F^2 m_b^5\over 192\pi^3}
\left|V_{cb}\right|^2\left[R\left({m_c^2\over m_b^2}\right)+
{\alpha_s\over\pi}A_0\left({m_c\over m_b}\right)\right]
\end{equation}
where ($r = m_c^2/m_b^2$)
\begin{equation}
R\left(r\right)=1-8r+8r^3-r^4-12r^2\log r
\end{equation}
is the leading order term in Eq.~(\ref{incform}). The function $A_0$
is given explicitly in~\cite{FLS}. Expanding $A_0$ and $R$ in the SV
limit gives
\begin{eqnarray}\label{a0exp}
A_0 &=& -{64\over 5} \left(\delta m\right)^5 +{96\over 5}
{\left(\delta m\right)^6\over m_b} \nonumber \\
&&+\left(-{632608\over33075} + {2048\over315} \log \left({2\delta
m\over m_b}\right)\right) {\left(\delta m\right)^7\over m_b^2},\\
\noalign{\smallskip} R &=& {64\over 5} \left(\delta m\right)^5 -
{96\over 5} {\left(\delta m\right)^6\over m_b}+ {288\over35}
{\left(\delta m\right)^7\over m_b^2}. \label{Rexp}
\end{eqnarray}

The terms in the expansion Eq.~(\ref{a0exp}) of $A_0$ must be obtained
from the $\alpha_s$ corrections to the exclusive form factors, using
the same mass renormalization scheme (in this case, on-shell
renormalization).  In HQET, the $\alpha_s$ corrections in the SV limit
can be computed by matching from QCD to the effective theory with
heavy $b$ and $c$ quarks at the scale $m_b$, which keeps the entire
functional dependence on $m_c/m_b$, followed by renormalization group
scaling from $m_b$ to some hadronic scale $\mu$ of order $\Lambda_{\rm
QCD}$. The Isgur-Wise function is the matrix element of HQET operators
renormalized at the scale $\mu$, and does not contain any large
logarithms.  The form factors to order $\alpha_s$ are given in
Eq.~(\ref{alpha-ff}).

The total $B$ decay rate at order $\left(\delta m\right)^5$ and
$\left(\delta m\right)^6$ is obtained by using the form factors
Eq.~(\ref{alpha-ff}) and substituting in
Eq.~(\ref{Ldexp},\ref{Ldsexp}) ignoring for the moment, the scale
factor $S$.  The result is \widetext
\begin{eqnarray}
&&\Gamma^L = {G_F^2\over 192\pi^3} \left|V_{cb}\right|^2\Biggl[
\left(1-{\alpha_s\over\pi}\right)\left({64\over 5} \left(\delta
M\right)^5 -{96\over 5}{\left(\delta M\right)^6\over M_B}\right) +
64\left(1- {4\alpha_s\over3\pi}\right) {\left(\delta M\right)^4 G\over
M_B} \nonumber\\ &&+{1\over M_B^2}\left(\delta
M\right)^7\left({352\over35} -{256\over35} \rho^2- {\alpha_s
\over\pi}\left({304\over315} - {704\over105}\rho^2\right)
+\ldots\right) \Biggr]+ \ldots, \label{eq:1}
\end{eqnarray}
%
where we have only retained terms of order $\left(\delta
M\right)^7$ in the $1/m^2$ corrections. The result Eq.~(\ref{eq:1}) is
precisely what is required for the first two terms of the inclusive
and exclusive decay rates to match at order $\alpha_s$. The terms of
order $\left(\delta M\right)^5$ and $\left(\delta M\right)^6$ agree.
The agreement between the $G \left(\delta M\right)^4$ terms can not be
checked, since the $\alpha_s$ corrections to the inclusive decays
including $1/m$ corrections have not been computed.

The $\left(\delta m\right)^7 \log\left(\delta m\right)$ term arises
from renormalization group scaling between $m_b$ and some
renormalization point $\mu$ which is of order $\Lambda_{\rm QCD}$. The
anomalous dimension vanishes as $w\rightarrow 1$, which is why the
$\log$ first appears multiplying the $\left(\delta m\right)^7$
term. All the form factors are multiplied by an overall scaling
factor~\cite{FGGW},
\begin{equation}\label{scalef}
S=\left({\alpha_s\left(m_b\right)\over
\alpha_s\left(\mu\right)}\right)^{\gamma\left(w\right)/2 b_0}
\end{equation}
where $\gamma\left(w\right)$ is the velocity dependent anomalous
dimension,
\begin{equation}\label{vdep}
\gamma\left(w\right)={16\over3}\left[
w{\log\left(w+\sqrt{w^2-1}\right) \over \sqrt{w^2-1}} -1\right] ,
\end{equation}
and $b_0$ is the first term in the QCD $\beta$-function. Expanding
Eq.~(\ref{scalef}) and Eq.~(\ref{vdep}) in $\alpha_s$ gives
\begin{eqnarray}\label{scaletwo}
S &=& 1+ \gamma\left(w\right){\alpha_s\over4\pi} \log{\mu\over
m_b}+{\cal O}\left(\alpha_s\right)^2 \nonumber\\ &=&1 + {8\over9}
{\alpha_s\over\pi}\left(w-1\right) \log{\mu\over m_b} +{\cal
O}\left(w-1\right)^2 .
\end{eqnarray}
The net effect of $S$ is to change the slope of the form factors at
zero recoil by a large logarithm. Having extracted this large log, we
refrain from denoting the scale at which $\alpha_s$ should be
evaluated, because the difference is of order $\alpha_s^2$.  The
contribution of the renormalization group scale factor $S$ to the
total rate can be obtained from Eq.~(\ref{Lexc}) by the substitution
\begin{equation}
\rho^2 \rightarrow \rho^2 - {8\alpha_s\over9\pi}\log{\mu\over m_b},
\end{equation}
which gives a contribution to the total rate of
\begin{equation}
{G_F^2\over 192\pi^3} \left|V_{cb}\right|^2\
{2048\over315}{\alpha_s\over\pi} {\left(\delta m\right)^7\over
m_b^2}\log{\mu\over m_b}.
\end{equation}
The $\log m_b$ dependence matches that in Eq.~(\ref{a0exp}).  The
remaining terms in Eq.~(\ref{a0exp}) are of order $\left(\delta
m\right)^7$, or of order $\left(\delta m\right)^7 \log \delta
m/\mu$. These will be discussed in the next subsection.

The order $\alpha_s$ corrections to the vector current decay have not
been calculated previously. The exclusive decay calculation gives a prediction
for the $\alpha_s$ corrections to the inclusive decay rate. Since the
exclusive decay rate to order $\left(\delta m\right)^7$ is given
entirely by $B\rightarrow D$, and $f_+\left(1\right)$ gets a
renormalization group scaling correction, but no finite $\alpha_s$
correction, one obtains
\begin{eqnarray}
\Gamma^V &=& {G_F^2 \over 192\pi^3}
\left|V_{cb}\right|^2\Biggl[{64\over5}\left(\delta m\right)^5 -
{96\over5}{\left(\delta m\right)^6\over m_b}\nonumber\\
&&+{512\over63}{\alpha_s\over\pi} {\left(\delta m\right)^7\over
m_b^2}\log{2\delta m\over m_b}+ {\cal O}\left(\delta
m\right)^7\Biggr].
\end{eqnarray}
We have verified the first two terms by a direct calculation of the
inclusive rate.

\subsection{The Differential Distributions}

The differential distributions for the exclusive and inclusive decays
agree to order $\alpha_s/m^2$. The $\alpha_s$ corrections to the
exclusive decay tensors are simply obtained by using the expressions
Eq.~(\ref{alpha-ff}) for the $\alpha_s$ corrections to the form
factors, multiplying by the scale factor Eq.~(\ref{scalef}) and
substituting in the full expressions Eq.~(\ref{wd1}--\ref{wds5}).

\begin{figure}
\moveright1in\hbox{\epsffile{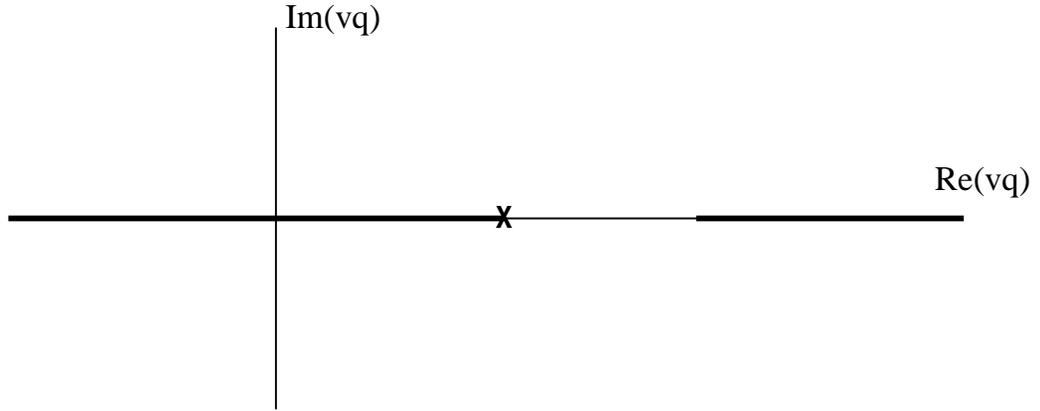}}
\caption{Analytic structure of $T_i$ in the complex $v\cdot q$
plane. The physical cut for semileptonic decay is along the real axis
for $q\cdot v \ge 0$. The free quark decay tensors have a pole marked
by a $\times$. The gluon bremsstrahlung graphs produce a cut to the
left of the pole. The left hand cut for $q\cdot v < 0$, and the right
hand cut correspond to crossed processes, and are not relevant for
semileptonic $b \to c$ decay.}
\label{fig:qvcut}
\end{figure}

The $\alpha_s$ corrections to the inclusive tensors arise from virtual
gluon corrections, and from gluon bremsstrahlung. The virtual and real
contributions are separately infrared divergent, but their sum is
finite. The virtual graph contributions to $W_i$ are proportional to
$\delta\left(\Delta_q\right)$, since the final state is still a single
quark. Thus the residue of the pole in Fig.~\ref{fig:qvcut} is
modified at order $\alpha_s$. The bremsstrahlung graphs have a gluon
and quark in the final state, so they have a two particle cut for $0
\le q\cdot v \le \delta m$, as show in Fig.~\ref{fig:qvcut}.  For the
exclusive and inclusive decay tensors to match, it must be that the
two particle cut is suppressed by $\left(\delta m/m_b\right)^2$
relative to the $\alpha_s$ corrections to the residue at the pole,
since the exclusive $D$ and $D^*$ states produce only a pole
contribution, and the excited states which produce a cut in the
exclusive decay tensors are of order $\left(\delta
m/m_b\right)^2$. This can be verified explicitly. Falk, Luke and
Savage have calculated the $\alpha_s$ corrections to the hadronic mass
distribution in Ref.~\cite{FLS}. Expanding their Eq.~(3.8) in the SV
limit gives (for $s\not=m_c^2$)
\begin{eqnarray}\label{flsresult}
{d\Gamma\over ds} &=&{G_F^2 \over 192\pi^3} \left|V_{cb}\right|^2
 {\alpha_s\over\pi}{1\over
s-m_c^2} {8\left(m_b^2-s\right)^5 \over 315 m_b^{9}} \Bigl[9
\left(m_b^2-s\right)^2 \nonumber\\ &&- 35m_b\, \delta m
\left(m_b^2-s\right) + 42m_b^2\, \delta m^2 \Bigr],
\end{eqnarray}
where $s$ is the invariant mass of the final hadronic system. We have
taken $s-m_b^2$ as of order $m_b\, \delta m$ and retained the leading
term in $ \delta m$, but kept the explicit singularity at $m_c^2$
since we intend to study the behavior about that singularity.
Eq.~(\ref{flsresult}) shows immediately that the two-particle cut's
contribution to $d\Gamma/ds$ is of order $\left(\delta m \right)^6$,
and to $\Gamma$ is naively of order $\left(\delta m
\right)^7$. However, there is a divergence as $s\rightarrow m_c^2$, so
that the total width due to the bremsstrahlung graphs has a
logarithmic contribution,
\begin{equation}
{G_F^2 \over 192\pi^3} \left|V_{cb}\right|^2 {\alpha_s \over \pi}
{2048\over315}
{\left(\delta m\right)^7 \over m_b^2} \log {m_b^2-m_c^2\over\mu^2}
\end{equation}
on integrating over the allowed region $m_c^2\le s \le m_b^2$, where
$\mu$ is an infrared regulator.  The infrared divergence is canceled
by the virtual graphs, which have a logarithmic contribution of the
form $-(2048/315) \left(\delta m\right)^7 \log m_b^2/\mu^2$. The sum
of the real and virtual graphs gives the infrared finite contribution
$(2048/315)\left(\delta m\right)^7\log(2\delta m/m_b)$ in
Eq.~(\ref{a0exp}). On the exclusive side, there are no infrared
divergences because of confinement effects. $B\rightarrow D,D^*$
decays give a contribution of order $-(2048/315) \left(\delta
M\right)^7 \log M_B^2/\Lambda^2$ to the decay width, and the excited
states give a contribution of order $(2048/315) \left(\delta
M\right)^7 \log 2M_B \delta M/\Lambda^2$, where $\Lambda$ is some
hadronic mass scale of order $\Lambda_{\rm QCD}$.

We have shown that the only corrections to the inclusive tensors $W_i$
at order $\left(\delta m\right)^5$ and $\left(\delta m\right)^6$ are
due to the virtual graphs, and are proportional to
$\delta\left(\Delta_q\right)$. Following the arguments of
sec.~\ref{sec:diff}, all that remains is to show that the coefficients
of the delta functions agree for the inclusive and exclusive
decays. The $\alpha_s$ corrections to the tensors are not known
explicitly, but we do know that their integrals over phase space
agree. We expect that the tensors themselves will agree to this order
as well.

\section{$1/m^2$ Corrections and Sum Rules}\label{sec:1/m2}

We have demonstrated consistency between the inclusive and exclusive
decay rates at order $1/m$. Excited states first contribute to the
decay rates at order $1/m^2$. Since the contribution of the excited
states to the decay rates are positive, one can get inequalities by
comparing the $1/m^2$ corrections to the inclusive and exclusive decay
widths.

\subsection{An Inequality on the Slope of the Isgur-Wise Function}

The inclusive $V-A$ decay width Eqs~(\ref{Linc2}) must be greater than
the sum of the $B\rightarrow D$ and $B\rightarrow D^*$ widths given in
Eq.~(\ref{Lexc}). The leading order term and the first correction
cancel, so one obtains an inequality on the $1/m^2$ terms. In the SV
limit, the $1/m^2$ terms can be organized as an expansion in powers of
$\Lambda_{\rm QCD}/\delta M$, since we are assuming that $m_b,m_c \gg
\delta M \gg \Lambda_{\rm QCD}$.  Keeping only the leading order terms
in the inequality (including $\alpha_s$ corrections), gives
\begin{equation}\label{rhobound}
\rho^2(\mu) > {1\over 4} + {\alpha_s \over \pi}\left({2563 \over 945}
 - {8\over 9} \log {2 \delta m \over \mu} \right).
\end{equation}
This is the Bjorken bound~\cite{bj}, $\rho^2 > 1/4$, including
$\alpha_s$ corrections.

Choosing $\mu=2\delta m$ in Eq.~(\ref{rhobound}) gives the inequality
\begin{equation}\label{rhobound2}
\rho^2(\mu) > {1\over 4} + {\alpha_s \over \pi}{2563 \over 945},
\end{equation}
which holds provided that $\mu$ is large compared with $\Lambda_{\rm
QCD}$, so that $\Lambda_{\rm QCD}/\mu$ terms can be neglected. For
example, if $\alpha_s=0.22$, its value at $m_b$, the bound is
\begin{equation}\label{nbound}
\rho^2(\mu) > 0.44.
\end{equation}
The bound Eq.~(\ref{rhobound2}) has no heavy quark mass dependence,
since $\rho^2(\mu)$ is a parameter of the effective theory.  The bound
was derived in the SV limit, but since $\rho^2$ is independent of
$m_b$ and $m_c$, it is valid for any numerical values of the heavy
quark masses, as long as $m_b, m_c \gg \Lambda_{\rm QCD}$.

The dependence of $\rho^2(\mu)$ for large values of $\mu$ is governed
by perturbation theory, and is known to be $\rho^2(\mu) =
\rho^2(\mu_0) + 8 \alpha_s/(9\pi) \log \mu/\mu_0$. Since $\rho^2(\mu)$
increases with $\mu$ the most restrictive bound on $\rho^2$ is
obtained by choosing $\mu$ in Eq.~(\ref{rhobound2}) to be as small as
possible, subject to the condition that $\Lambda_{\rm QCD}/\mu \ll 1$.

The quantity $\rho^2(\mu)$ is not directly measurable, but it can be
related to the slope of the decay form factors, which can be
measured. It is convenient to define ${\cal
F}(w)=\left(2/(1+w)\right)\left(f_0(w)/f_0(1)\right)$, which is equal
to the Isgur-Wise function $\xi(w)$ when $1/m$ and $\alpha_s$
corrections are neglected.  Including the $\alpha_s$ corrections
computed in Ref.~\cite{FG,neubert} gives
\begin{equation}
{d {\cal F}\over dw}(1) \approx -\rho^2 - 0.01,
\end{equation}
where we have used $\mu=m_c$ and $m_c/m_b=0.3$ to evaluate the
correction.  Thus the slope of ${\cal F}$ at $w=1$ is bounded above by
$-0.45$.

The ALEPH collaboration~\cite{aleph} quotes a slope of $-0.39 \pm
0.21$, while the CLEO collaboration~\cite{cleo} quotes a slope of
$-0.84 \pm 0.13$.  Both used linear fits; a QCD derived
parametrization brings the ALEPH slope into closer agreement with the
above CLEO slope~\cite{blg}. All are consistent with the
bound~(\ref{nbound}).

The inequality Eq.~(\ref{rhobound}) depends on $\delta m$.  The
analysis of the previous section indicates that the effect of the
excited states is to replace $\delta m$ in the logarithm by a scale of
order $\Lambda_{\rm QCD}$. This would further strengthen the bound on
the slope of the form factor at zero recoil.

\subsection{Inequalities on Hadronic Matrix Elements}\label{sec:kbound}

The inequality,
\begin{equation}\label{KGineq}
K+G > 0
\end{equation}
has been obtained in Ref.~\cite{BSUV2} by studying $W_{ii}$ for vector
current decay. Let us here consider the integral of $W_i^i$ (summed on
$i$) over the allowed kinematic region for semileptonic decay. This
can be thought of as the decay rate for a $B$ hadron in a modified
theory, where $W_i^i$ replaces the tensor combination in
Eq.~(\ref{diffrate}).  The positivity conditions of a ``real'' decay
must still hold, since $W_i^i$ is the sum of the absolute value
squared of hadronic matrix elements.  The advantage of working with
$W_i^i$ is that the matrix element $K$ appears in the decay rate at
leading order in $1/m$. The reason is that the $D$ and $D^*$ rates
vanish at zero recoil, and so are suppressed over the entire kinematic
region in the SV limit.

The inclusive rate is given straightforwardly by integrating
Eqs.~(\ref{teqns}) using Eq.~(\ref{diffrate}) with the square brackets
replaced by $3W_1+ (v\cdot q^2 - q^2) W_4$,
\begin{equation}\label{incsum}
\int W_i^i = {G_F^2\over192\pi^3} {16 (\delta m)^3 \over
5m_b^2}\left[9(\delta m)^2 + 40 K + {\cal O}(1/m_b^3)\right].
\end{equation}
The exclusive rate is computed from Eqs.~(\ref{wd1})--(\ref{wds5}),
\begin{equation}\label{exsum}
\int W_i^i = {G_F^2\over 192\pi^3}{(\delta M)^5 \over M_B^2}{144\over
       5} + exc + {\cal O}(1/M_B^3),
\end{equation}
where $exc$ denotes the contribution from excited states, and is
positive.  Comparing the two rates leads to the inequality
\begin{equation}\label{Kbound}
K > 0.
\end{equation}
Though the derivation of Eq.~(\ref{Kbound}) depends on the SV limit,
it is valid even away from this limit, since $K$ is defined in the
effective theory and is independent of the quark masses.  A closely
related sum rule has been analyzed under different assumptions in
Ref.~\cite{BSUV2}. Our relation Eq.~(\ref{Kbound}) agrees with
Eq.~(121) of that reference when $m_b = m_c$. The inequality
Eq.~(\ref{KGineq}) was obtained in Ref.~\cite{BSUV2} by considering
$b\rightarrow c$ decays, and then taking the limit $m_c \gg m_b$. We
obviously can not take this limit here, since we are considering a
physical decay.

There is an important correction to Eq.~(\ref{Kbound}) which
invalidates the above derivation of the inequality. As pointed out by
Falk, Luke and Savage,
there are $\alpha_s$ corrections which can not be neglected in the
derivation of Eq.~(\ref{Kbound})~\cite{meluke}. These corrections are of
order $\alpha_s \left(\delta m\right)^2$, which are parametrically
much larger than $K$, which is of order $\Lambda_{\rm QCD}^2$. It is
straightforward to check that the sign of the corrections is such that
any value of $K$ is allowed.

\section{$\Lambda_b \to \Lambda_c$ Decay}\label{sec:lambda}

The entire analysis of $B\to D,D^*$ decay presented so far can be
extended to $\Lambda_b\to \Lambda_c$ decays. We will not present the
details of all the computations here, since the results are very
similar to those for $B$ meson decay.

The form factors for $\Lambda_b \to \Lambda_c$ decay are defined
by~\cite{lambdas}
\begin{eqnarray}
\left\langle \Lambda_c,v^\prime\right| V^\mu
\left|\Lambda_b,v\right\rangle &=& \bar
u\left(v^\prime\right)\left(F_1\gamma^\mu+F_2v^\mu+F_3v^{\prime \mu}
\right)u\left(v\right), \nonumber\\ \left\langle
\Lambda_c,v^\prime\right| A^\mu \left|\Lambda_b,v\right\rangle &=&
\bar u\left(v^\prime\right)\left(G_1\gamma^\mu+G_2v^\mu+G_3v^{\prime
\mu} \right)\gamma_5 u\left(v\right),\nonumber
\end{eqnarray}
where the states are normalized to $v^0$.  The form factors for
$\Lambda_b \to \Lambda_c$ decay near zero recoil can be computed using
HQET. The expansion of the form factors around $w=1$ is given in
Eq.~(\ref{lamffexp}).

The decay tensors $W_i$ for $\Lambda_b \to \Lambda_c$ are computed by
squaring the matrix element and summing over the polarizations of the
final state $\Lambda_c$. The results for the spin-independent decay
tensors are summarized in Eq.~(\ref{lamw}).  Decay tensors $S_i$ for
inclusive polarized $\Lambda_b$ decay were given in
Ref.~\cite{MW}. The decay tensors for $\Lambda_b \to \Lambda_c$ and
$\Lambda_b \to X_c$ agree in such a way that the differential decay
rates are equal to two orders in $1/m$ for the spin-dependent and
spin-independent terms, i.e. $W_{1,2}$ and $S_{6,8}$ agree to two
orders in $1/m$, while $W_{3,5}$ and $S_{1,2,7,9}$ agree to first
order in $1/m$. It is straightforward to verify this result, and the
details are omitted here.

The total decay rate for $\Lambda_b\rightarrow \Lambda_c$ decay for a
left handed current is
\begin{eqnarray}
&&\Gamma^L\left(\Lambda_b\to \Lambda_c\right) = {G_F^2\over 192\pi^3}
\left|V_{cb}\right|^2\Biggl[ {16\over 5} \left(\delta
M\right)^5\left[\left(F_1+F_2+F_3\right)^2 + 3 G_1^2\right] -{8\over
5}{\left(\delta M\right)^6\over M_{\Lambda_b}}
\left[\left(F_1+F_2+F_3\right)\left(3F_1+5F_2+F_3\right)
\right.\nonumber\\ &&\left.+ G_1\left(9G_1+2G_2+2G_3\right)\right]
+{4\over 35 }{\left(\delta M\right)^7\over M_{\Lambda_b}^2} \left[
20\left(F_1+F_2+F_3\right)\left({dF_1\over dw}+{dF_2\over
dw}+{dF_3\over dw} \right) \right.+ 44 G_1 {dG_1\over dw} + 24 F_1^2 +
78 F_1 F_2 \nonumber\\ &&+ 65 F_2^2 + 22 F_1 F_3 + 46 F_2 F_3 + 9
F_3^2 + 48 G_1^2 + 42 G_1 G_2 + 5 G_2^2 + 14 G_1 G_3 + 10 G_2 G_3 + 5
G_3^2\Biggr] +\ldots, \label{eq:12}
\end{eqnarray}
where $\delta M=M_{\Lambda_b}-M_{\Lambda_c}$ and all the form factors
are evaluated at zero recoil.

The rate for $\Lambda_b\rightarrow \Lambda_c$ decay including
$\alpha_s$ corrections is obtained by using the form factors
Eq.~(\ref{lamffexp}), and substituting into Eq.~(\ref{eq:12}),
\begin{eqnarray}
&&\Gamma^L\left(\Lambda_c\right) = {G_F^2\over 192\pi^3}
\left|V_{cb}\right|^2\Biggl[ \left(1-{\alpha_s\over\pi}\right)
\left({64\over 5} \left(\delta M\right)^5 -{96\over 5}{\left(\delta
M\right)^6\over M_{\Lambda_b}}\right) \nonumber\\ &&+{1\over
M_{\Lambda_b}^2} \left(\delta M\right)^7\left({288\over35}
{}-{256\over35} \rho^2 + {\alpha_s \over\pi}\left({32\over45} +
{704\over105}\rho^2\right) \right) + {2048\over315} \log
{m_b\over\mu}\Biggr] + \ldots, \label{eq:2}
\end{eqnarray}
There are no $G$ terms in the $\Lambda_b$ decay rate because the light
degrees of freedom have spin zero, so matrix elements of the $\bar b
\sigma^{\mu\nu}G_{\mu\nu}b$ operator vanish in the $\Lambda_b$. Also
note that there are no $\bar\Lambda$ terms in Eq.~(\ref{eq:2}), even
though the individual form factors depend on $\bar\Lambda$ through
$\epsilon_{b,c}$.

Comparing Eq.~(\ref{eq:2}) with the inclusive decay rates
Eq.~(\ref{incalpha}) gives the bound on the slope of the Isgur-Wise
function for baryons,
\begin{equation}\label{rhoboundb}
\rho^2(\mu) > {\alpha_s \over \pi}\left({2563 \over 945} - {8\over 9}
\log {2 \delta m \over \mu} \right),
\end{equation}
which gives the $\alpha_s$ correction to the lowest order result
$\rho^2\ge 0$\cite{IWY}.  An analysis similar to that for mesons gives
\begin{equation}\label{sbound2}
\rho^2(\mu) > {\alpha_s \over \pi}{2563 \over 945},
\end{equation}
valid for $\mu$ large enough that $\Lambda_{\rm QCD}/\mu \ll 1$.  For
$\alpha_s=0.22$, the bound is $\rho^2 > 0.19$. The bound for the
physical form factor $G_1$ is that the logarithmic slope is bounded
above by $-0.20$, using the relation between the Isgur-Wise function
and the form factors given in~\cite{FG,neubert}, $\mu=m_c$, and
$m_c/m_b=0.3$.

\section{Conclusions}

We have shown by explicit computation that the decay widths for $B\to
X_c l \bar \nu$ and $B\to \left(D+D^*\right) l \bar \nu$ decays agree
to two orders in the $1/m$ expansion in the SV limit, including
$\alpha_s$ radiative corrections. The agreement is non-trivial because
the $1/m$ term depends on the non-perturbative matrix element $G$, and
because the parton decay at order $\alpha_s$ involves the two-body
hadronic final state $c g$. The matrix element $G$ from the operator
product expansion enters in precisely the way necessary to compensate
for the difference between quark and meson kinematics. The two-body
charm-glue cut enters at $1/m^2$, as necessary to match the continuum
contribution from multi-hadron states.  The differential rates also
agree to two orders in $1/m$. The electron spectra agree to two orders
in $1/m$ without any smearing over $E_e$.  The double differential
rates agree as well, provided one smears over a kinematic region at
least as large as $\Lambda_{QCD}$. These results show that (smeared)
local duality holds in our example to two orders in $1/m$. This is a
non-trivial check on the validity of local duality, since
non-perturbative corrections come in at first order in $1/m$.

The partial width for $B$ decays to excited states is positive, so one
obtains an inequality on the $1/m^2$ corrections to the decay widths
which gives the $\alpha_s$ corrections to Bjorken's inequality on
$\rho^2$. A conservative estimate of the appropriate hadronic scale
leads to an upper bound on the slope of a measurable decay form factor
of $-0.45$. A similar analysis leads to the conclusion that the
kinetic energy operator of the heavy quark has positive matrix
element, $K > 0$, if one neglects $\alpha_s$ corrections. Unlike the
case of the Bjorken bound, the $\alpha_s$ terms are not merely a
correction to the leading order result and invalidate the
derivation of the $K>0$ bound.

We have also shown that the differential decay distributions and total
widths for polarized and unpolarized $\Lambda_b \to X_c l \bar \nu$
and $\Lambda_b \to \Lambda_c l \bar \nu$ agree to two orders in $1/m$,
in the SV limit. At order $1/m^2$ we find $\alpha_s$ corrections to
Bjorken's inequality on the slope of the baryon form factor.  The
logarithmic slope of the form factor $G_1$ has an upper bound of
$-0.20$.

\acknowledgments

We would like to thank M.B.~Wise for several useful discussions. We
would like to thank A.F.~Falk, M.E.~Luke and M.J.~Savage for
discussions on the results of Ref.~\cite{FLS}, and for pointing out
that $\alpha_s$ corrections are important for the results in
Sec.~\ref{sec:kbound}.

This work was supported in part by the Department of Energy under
Grant No.  DOE-FG03-90ER40546. B.G. was supported in part by a grant
from the Alfred P.~Sloan Foundation.  A.M. was supported in part by
the PYI program, through Grant No.~PHY-8958081 from the National
Science Foundation.

\widetext

\appendix

\section{Hadronic Tensors For Inclusive Decay}\label{app:tensors}

The tensors $T_i$ for semileptonic $B$ decay are summarized here. The
results are from Ref.~\cite{BKSV}, and have been converted to the
notation and sign conventions used in Ref.~\cite{MW}. The tensors for
unpolarized $\Lambda_b$ decay are obtained by setting $G=0$, and
replacing $K$ in Eq.~(\ref{kdef}) by the corresponding matrix element
in the $\Lambda_b$.
\begin{mathletters}\label{teqns}
\begin{eqnarray}
&&T_1^{VV}={1\over \Delta_q} \left[\left(m_b-m_c-q\cdot v\right)+
\left(G+K\right){1\over m_b}\left({1\over3} + {m_c\over
m_b}\right)\right] \\ &&+ {2\over \Delta_q^2 m_b} \left[-{1\over3}
G\left[\left( 4 m_b-3q\cdot v\right)\left(m_b-m_c-q\cdot v\right)+ 2
\left(q\cdot v^2 - q^2\right)\right] + K\left[q\cdot
v\left(m_b-m_c-q\cdot v\right)- {2\over 3} \left(q\cdot v^2 -
q^2\right)\right] \right]\\ &&+{8\over 3\Delta_q^3} K
\left(m_b-m_c-q\cdot v\right) \left(q\cdot v^2 - q^2\right),\\
\noalign{\medskip} &&T_2^{VV}={1\over \Delta_q}
\left[2m_b+{10\over3m_b} \left(G+K\right)\right] + {4\over
3\Delta_q^2} \left[-2G\left( m_b-m_c\right)+ 5 G q\cdot v + 7 K q\cdot
v\right] +{16\over 3\Delta_q^3}m_b K\left(q\cdot v^2 - q^2\right),\\
\noalign{\medskip} &&T_4^{VV} = {8\over 3 m_b
\Delta_q^2}\left(K+G\right),\\ \noalign{\medskip} &&T_5^{VV} =
{}-{1\over \Delta_q} - {2\over3\Delta_q^2}\left[5 {q\cdot v\over m_b}
\left(G+K\right) + 4 K\right] - {8\over 3\Delta_q^3}K\left(q\cdot v^2
{}- q^2\right),\\ \noalign{\medskip} &&T_3^{AV} = -{1\over \Delta_q} -
\left[-4 G + {10\over 3} \left(K+G\right){q\cdot v\over
m_b}\right]{1\over\Delta_q^2} -{8\over 3\Delta_q^3}K\left(q\cdot v^2 -
q^2\right),\\ \noalign{\medskip} &&T_3^{VV} =
T_1^{AV}=T_2^{AV}=T_4^{AV}=T_5^{AV}=0,
\end{eqnarray}
\end{mathletters}
and $T_i^{AA}$ is obtained from $T_i^{VV}$ by changing the sign of
$m_c$. Here $\Delta_q = m_b^2 - m_c^2 - 2 m_b q\cdot v + q^2$.

\section{$B\rightarrow D,D^*$ Form Factors}\label{app:ffalphas}
The expansions of the $B\rightarrow D$ and $B\rightarrow D^*$ form
factors around $w=1$ (using the results in~\cite{FG,neubert}) are
summarized here. The deviation $w-1$ is treated as order $\left(\delta
m\right)^2$.
\begin{mathletters}\label{alpha-ff}
\begin{eqnarray}
f_+ &=& {M_B+M_D\over 2\sqrt{M_B M_D}}\left[1 + {5\alpha_s\over 18
\pi} \left({\delta m\over m_b}\right)^2 + f_+^{(2)}{1\over m_b^2}+
{\cal O}\left(\delta m\right)^3\right]S(w)\, \xi(w) ,\\
\noalign{\smallskip} f_0 &=& 2\sqrt{M_B M_{D^*}}\left[{1+w\over2} +
f_0^{(2)}{1\over m_b^2}+
{\alpha_s\over\pi}\left(-{2\over3}+{2\over9}\left(w-1\right) +
{1\over6}\left({\delta m\over m_b}\right)^2\right)+ {\cal
O}\left(\delta m\right)^3\right]S(w)\,\xi(w) ,\\ \noalign{\smallskip}
a_+ &=& -{1\over 2}\sqrt{{1\over M_B M_{D^*}}}\left[1 +a_+^{(1)}
{1\over m_b} + {\alpha_s\over\pi}\left(-{2\over3} + {1\over9}{\delta
m\over m_b}\right)+ {\cal O}\left(\delta m\right)^2\right]S(w)\,
\xi(w) ,\\ \noalign{\smallskip} g &=& \sqrt{{1\over M_B
M_{D^*}}}\left[1 + {2\alpha_s\over 3 \pi}+{\cal O}\left(\delta
m\right)^2\right]S(w)\, \xi(w) .
\end{eqnarray}
\end{mathletters}
%
The Isgur-Wise function $\xi(w)$ can be expanded about
$w=1$,
\begin{equation}
\xi(w) = 1 - \rho^2 \left(w-1\right) + \ldots,
\end{equation}
which defines the slope parameter $\rho^2$. The renormalization group
scaling factor $S(w)$ is
\begin{equation}\label{app:scalef}
S=\left({\alpha_s\left(m_b\right)\over
\alpha_s\left(\mu\right)}\right)^{\gamma\left(w\right)/2 b_0}
\end{equation}
where $\gamma\left(w\right)$ is the velocity dependent anomalous
dimension,
\begin{equation}\label{app:vdep}
\gamma\left(w\right)={16\over3}\left[
w{\log\left(w+\sqrt{w^2-1}\right) \over \sqrt{w^2-1}} -1\right] ,
\end{equation}
and $b_0$ is the first term in the QCD $\beta$-function. Expanding
$\gamma\left(w\right)$ around $w=1$ gives
\begin{equation}
\gamma\left(w\right) = {32\over 9} \left(w-1\right) + \ldots.
\end{equation}

\section{$\Lambda_b$ Form Factors}

The form factors for $\Lambda_b\rightarrow \Lambda_c$ decay for a
left-handed current are
\begin{mathletters}\label{lamw}
\begin{eqnarray}
{4\over M_{\Lambda_c}} W_1 &=& X_1
\delta\left(\Delta_{\Lambda_c}\right),\\ {4\over M_{\Lambda_c}} W_2
&=& \left[X_2+2{M_{\Lambda_b}\over M_{\Lambda_c}} X_5 + X_4
\left({M_{\Lambda_b}\over M_{\Lambda_c}}\right)^2\right]
\delta\left(\Delta_{\Lambda_c}\right),\\ {4\over M_{\Lambda_c}} W_3
&=& {X_3\over M_{\Lambda_c}} \delta\left(\Delta_{\Lambda_b}\right), \\
{4\over M_{\Lambda_c}} W_4 &=& {X_4\over M_{\Lambda_c}^2}
\delta\left(\Delta_{\Lambda_c}\right),\\ {4\over M_{\Lambda_c}} W_5 &=
&-\left[{X_5\over M_{\Lambda_c}}+ {M_{\Lambda_b}\over
M_{\Lambda_c}^2}X_4\right]\delta\left(\Delta_{\Lambda_c}\right),
\end{eqnarray}
\end{mathletters}
where $X_i$ are defined by \widetext
\begin{mathletters}
\begin{eqnarray}
X_1 &=& \left(w-1\right)F_1^2 + \left(w+1\right) G_1^2,\\ X_2 &=&
\left(w+1\right) F_2^2 + 2 F_1 F_2 + \left(w-1\right) G_2^2 + 2 G_1
G_2, \\ X_3 &=& 2 F_1 G_1,\\ X_4 &=& \left(w+1\right) F_3^2 + 2 F_1
F_3 + \left(w-1\right) G_3^2 - 2 G_1 G_3, \\ X_5 &=& \left(w+1\right)
F_2 F_3 + F_1\left(F_1+F_2+F_3\right) + \left(w-1\right)G_2 G_3 +
G_1\left(G_1+G_3-G_2\right),
\end{eqnarray}
\end{mathletters}
and
\begin{equation}
\Delta_{\Lambda_c} = \left(p_{\Lambda_b}-q\right)^2 - M_{\Lambda_c}^2.
\end{equation}
The form factors for more general currents can be trivially obtained
from the above by rescaling the axial or vector current form factors.

The expansion of the $\Lambda_b$ form factors around $w=1$ at order
$\alpha_s$ is (using the results in~\cite{FG,neubert})
\begin{mathletters}\label{lamffexp}
\begin{eqnarray}
F_1 &=&
\left[\left(1+{2\alpha_s\over3\pi}\right)\left(1+\epsilon_b+\epsilon_c\right)
+ {\alpha_s\over\pi}\left({w-1\over9}+{\left(\delta m\right)^2\over 6
m_b^2} \right) + {\cal O}\left(\delta m\right)^3\right]S(w)\,
\zeta(w)\\ F_2 &=&
\left[-{\alpha_s\over3\pi}\left(1+\epsilon_b+\epsilon_c\right)
{}-\epsilon_c - {\alpha_s\over9\pi}\left(1+\epsilon_b+\epsilon_c\right)
{\delta m\over m_b} + {\alpha_s\over9\pi}\left({w-1}-{\left(\delta
m\right)^2\over 2 m_b^2} \right) + {\cal O}\left(\delta
m\right)^3\right]S(w)\, \zeta(w)\\ F_3 &=&
\left[-{\alpha_s\over3\pi}\left(1+\epsilon_b+\epsilon_c\right)
{}-\epsilon_b + {\alpha_s\over9\pi}\left(1+\epsilon_b+\epsilon_c\right)
{\delta m\over m_b} + {\alpha_s\over9\pi}\left({w-1}+{\left(\delta
m\right)^2\over 2 m_b^2} \right) + {\cal O}\left(\delta
m\right)^3\right]S(w)\, \zeta(w)\\ G_1 &=& \left[1 -
{2\alpha_s\over3\pi} + {\alpha_s\over\pi}\left({5\left(w-1\right)\over
9} +{\left(\delta m\right)^2\over 6 m_b^2}\right)+ {\cal
O}\left(\delta m\right)^3\right]S(w)\, \zeta(w), \\ G_2 &=& \left[
{}-{\alpha_s\over\pi}\left(1+2\epsilon_b\right)\left({7\over9} + {\delta
m\over 3 m_b}\right) - \epsilon_c\left(1+{8\alpha_s\over9\pi}\right)
+{\alpha_s\over 90\pi} \left(22\left(w-1\right) - 17 {\left(\delta
m\right)^2\over m_b^2}\right) + {\cal O}\left(\delta
m\right)^3\right]S(w)\,\zeta(w), \\ G_3 &=& \left[
{\alpha_s\over\pi}\left(1+2\epsilon_c\right)\left({7\over9} - {\delta
m\over 3 m_b}\right) +
\epsilon_b\left(1+{8\alpha_s\over9\pi}\right)-{\alpha_s\over 90\pi}
\left(22\left(w-1\right) +13 {\left(\delta m\right)^2\over
m_b^2}\right) + {\cal O}\left(\delta m\right)^3\right]S(w)\,\zeta(w).
\end{eqnarray}
\end{mathletters}
Here
\begin{equation}\label{epsb}
\epsilon_{b,c} = {\bar\Lambda\over2m_{b,c}},
\end{equation}
and the scale factor $S(w)$ is the same as for meson form factors.  In
the above expansion, we have omitted terms that arise from the
$1/m_b^2$ operators in HQET, analogous to the $f_+^{(2)}/m_b^2$ terms
in the meson form factors. Such terms do not affect the inequality on
$\rho^2$.  Note that $\bar\Lambda$ and $\rho^2=d\zeta/dw(1)$ for
baryons are different from $\bar\Lambda$ and $\rho^2=d\xi/dw(1)$ for
mesons, even though they are denoted by the same symbol.

\vfill\break\eject

\def\epsfsize#1#2{0.50#1}

\eject

\def\epsfsize#1#2{0.80#1}

\end{document}